\documentclass[onecolumn,
10pt,tightenlines,
notitlepage,
nofootinbib,
superscriptaddress,
amsfonts,amsmath]{revtex4}
\usepackage{amssymb}
\usepackage{epsfig}
\usepackage{calc}
\usepackage{graphicx}
\usepackage[ansinew]{inputenc}
\usepackage{multirow}
\usepackage{subcaption}
\usepackage{tensor}

\usepackage[dvipsnames]{xcolor}
\usepackage{natbib}

\DeclareMathOperator{\sign}{sign}

\begin{document}
\title{Rotating black holes embedded in a cosmological background for scalar-tensor theories}

\author{Eugeny Babichev}
\affiliation{Universit\'e Paris-Saclay, CNRS/IN2P3, IJCLab, 91405 Orsay, France}
\author{Christos Charmousis}
\affiliation{Universit\'e Paris-Saclay, CNRS/IN2P3, IJCLab, 91405 Orsay, France}
\author{Nicolas Lecoeur}
\affiliation{Universit\'e Paris-Saclay, CNRS/IN2P3, IJCLab, 91405 Orsay, France}

\begin{abstract}
We present solutions of  DHOST theories describing a rotating black hole embedded in an expanding universe. 
The solution is constructed by conformal transformation 
of a stealth Kerr(-de Sitter) black hole. 
The conformal factor depends explicitly on the scalar field -- but not on its derivative -- and defines the new theory.
The scalar field of the stealth Kerr(-de Sitter) solution depends on time, leading to the time-dependence of the obtained conformal metric, with cosmological asymptotics at large distances. 
We study the properties of the obtained metric by considering regular null geodesic congruences, and identify trapping black hole and cosmological horizons.
\end{abstract}
\maketitle

\section{Introduction}

The Schwarzschild metric  and Friedmann universe are probably the most simple and well-known non trivial solutions of General Relativity (GR). 
The former represents a static black hole, while the latter describes a homogeneous isotropic universe. 
There is very little doubt that black holes do exist, although, rather than being static they have rotation, which is entailed in the Kerr metric. 
On the other hand, the evolution of our Universe at large scales follows the Friedmann equations.
It is then rather tempting to find a solution that combines both a black hole and a homogeneous and isotropic universe. 
This task, apart from being an exciting theoretical challenge, has also an interesting practical purpose, e.g. it is important for understanding the formation and evolution of primordial black holes in the early universe.
For an exponentially  expanding universe, such a solution is well-known and given by the Schwarzschild-de Sitter (Schwarzschild-dS) metric. 
The relative simplicity of this solution, and in particular the existence of coordinates where the metric is static, can be accounted for the form of the matter. Indeed the cosmological term has a constant density during the evolution and hence does not accrete onto a black hole, thus keeping the mass of the black hole constant. 

For a more general  Friedmann-Lema\^{i}tre-Robertson-Walker (FLRW) asymptotic, McVittie proposed a metric, which was believed to describe a point-like object embedded in an FLRW universe~\cite{McVittie:1933zz}. 
It was however later understood that it cannot describe a point-like object, because of a singularity inherited in the solution. 
Likewise, the McVittie metric cannot describe a black hole in the universe, due to presence of the same singularity at the would-be event horizon{\footnote{
This particular problem is alleviated for McVittie spacetimes if late time cosmology is dominated by a positive cosmological constant~\cite{Kaloper:2010ec}.}}. 
Besides, the density and pressure of the McVittie fluid source are not related by an equation-of-state, indeed, for this solution the pressure is a function of not only the density, but also the coordinates. 
Physically the problems of the McVittie solution are related to postulating rather than finding the metric via the Einstein equations. 
In particular, a zero flow of fluid onto a black hole was assumed, which is only justified in the case of the cosmological constant.
Indeed, at least in the test fluid approximation black holes do accrete surrounding fluid~\cite{Bondi:1952ni}.
The Sultana-Dyer spacetime~\cite{Sultana:2005tp} is another attempt to construct a black hole in an FLRW universe. 
It has an asymptotic FLRW form with a scale factor corresponding to dust, while the matter source is two-fluid, one being a time-like dust and the other is radial null dust. 
Interestingly, the Sultana-Dyer metric is conformal to the Schwarzschild metric, and accretion takes place for this solution. 
Apart from the fact that one needs to assume two-fluid model, the solution contains other pathologies, and we refer the reader to~\cite{Faraoni:2015ula}, where other time-dependent metrics are also considered in detail. 
A more physical approach to constructing black hole solutions embedded in cosmological background is to solve directly Einstein equations using various approximations. For example one may assume the separation of black hole and cosmological horizon scales~\cite{Babichev:2018ubo}, or a slow roll in the case where the source is a scalar field~\cite{Chadburn:2013mta}.

It should be stressed that all the approaches mentioned above only attempt to find non-rotating black holes in an FLRW universe. To the best of our knowledge, the only example of a rotating black hole embedded in a cosmological background is the Kerr-de Sitter (Kerr-dS) solution first discovered by Carter~\cite{Carter:1968rr}. 
No extensions to more general asymptotic cosmological backgrounds were reported. 
Taking into account issues related to the construction of spherically symmetric black holes in FLRW, the task of finding rotating analogs embedded in general Friedmann universe in GR looks to be extremely difficult.

There are extra challenges when we consider such solutions in modified gravity. 
There is a strong motivation to study modified gravity which comes, in particular, from the unexplained nature of dark energy, non-detection of dark matter particles, as well as theoretical issues of GR. 
Moreover, the growing data of observations, such as gravitational waves from binaries~\cite{LIGOScientific:2017vwq}, star trajectories orbiting the supermassive black hole at the center of the Milky Way~\cite{GRAVITY:2018ofz}, the recent images of black holes by the Event Horizon Telescope~\cite{EventHorizonTelescope:2019dse}, provide us with increasingly precise knowledge about the properties of gravitating objects at various curvature scales and gravitational waves. 
It is then important to search and study gravity modifications, which satisfy the tight experimental constraints which give bench marks for testing GR as well as indicating possible deviations from GR.
The two most discussed directions of modified gravity in the recent decade were the massive (and bi-) gravity (see the review Ref.~\cite{deRham:2014zqa}) and the degenerate higher order scalar-tensor (DHOST) theory~\cite{Langlois:2015cwa,Crisostomi:2016czh}.
The problem of finding exact solutions in  modified theories of gravity is rooted in the extra degrees of freedom, e.g. an extra dynamical metric in bi-gravity or a propagating scalar in scalar-tensor theories. A simplifying hypothesis which helps to solve the field equations is the assumption of a GR background. Such solutions are sometimes referred to as stealth solutions. 
It has not been such a long time, since counterparts of rotating GR solutions were found, in particular in massive (bi)-gravity~\cite{Babichev:2014tfa} (see also~\cite{Ayon-Beato:2015qtt} for the Kerr-dS extension), and in scalar-tensor theory~\cite{Charmousis:2019vnf}. 
In all these solutions the (physical) metric has the same form as in GR, i.e. the form of the Kerr(-dS) metric. 
A straightforward extension to find a non-GR background in modified gravity thus seems quite a futile task, taking into account the complexity of the field equations. 

DHOST theory, however, inherits a crucial property that allows to search for new solutions in a different manner.
Indeed, DHOST theory is stable under conformal and disformal transformations with generic coefficients which may depend on the scalar field and/or its kinetic term~\cite{Achour:2016rkg}.
More precisely, such a transformation takes one DHOST type Ia theory\footnote{DHOST theory has various subclasses, among which type Ia is the only phenomenologically viable, see~\cite{Langlois:2018dxi}.} 
to another theory of the same type. 
Therefore, once having a solution in a particular theory, via conformal/disformal transformation one generates another solution in a different theory. 
This trick has been used, in particular in~\cite{Babichev:2017lmw} to transform the solution of Schwarzschild-dS in a theory with speed of graviton different from one, to the same solution (with rescaled mass and cosmological constant) in a theory in which the graviton and light speed coincide. 
The metric after the transformation takes again the GR form, therefore no new solution of the metric is generated in the end. 
However, the use of disformal transformation to the solution of~\cite{Charmousis:2019vnf}, where the metric has a Kerr(-dS) form, gives an interesting non-trivial stationary black hole solution with non-GR metric, and properties quite distinct from the Kerr solution~\cite{Anson:2020trg} (see also~\cite{BenAchour:2020fgy} for studies of properties of the obtained metric). 

In this paper we use the same technique to generate new solutions. 
Instead of making use of disformal transformations, we apply a simple conformal transformation to the black hole solution with non-trivial scalar hair and the Kerr(-dS) metric of Ref.~\cite{Charmousis:2019vnf}. 
The conformal parameter depends only on the scalar field, but not on its derivatives. This allows us to generate new exact solutions of black holes embedded in a Friedmann universe. They are solutions of  DHOST theories. Since we start from a seed rotating metric, the resulting solution describes a rotating black hole with increasing mass. 
The plan of the paper is as follows: in Sec. II, we study the conformal transformation of a seed Kerr metric, presenting the obtained solution and characterizing it in detail as a cosmological black hole by the formalism of trapping horizons; while in Sec. III, we turn on to the conformal transformation of a seed Kerr-dS metric. We then present our conclusions in the last section.

\section{Cosmological Kerr black holes as vacuum solutions of DHOST theory}

The main steps in the construction are summed up in the following : we start by considering a known solution of a DHOST theory with flat (or dS) asymptotics for the metric $g_{\mu\nu}$ and a non-trivial time-like scalar field $\phi$. Then by use of a conformal transformation with conformal factor $C(\phi)$, we transform this seed metric $g_{\mu\nu}$ to a new metric $C(\phi)g_{\mu\nu}$, which is again a solution of (a different) DHOST theory. 
Since $\phi$ is time-like, we choose it to be (proportional to) a time coordinate, so that the new metric $C(\phi)g_{\mu\nu}$ asymptotically has an FLRW conformal form $C(\phi)g_{\mu\nu}^\text{(flat)}$ with $\phi$ being the conformal time\footnote{This is the case when the seed metric $g_{\mu\nu}$ is asymptotically flat. The case where it is asymptotically dS is slightly more subtle and will be treated in Sec. III.}. The black hole coordinates are also accordingly adapted to this time coordinate. 
Note that the causal structure is preserved under a conformal transformation, implying e.g. that if the seed solution is taken in a theory which has the speed of graviton equals to the speed of light, then in the resulting theory gravity propagates also with the speed of light. 
In this section we consider only the asymptotically Minkowski solutions as a seed metric, while the asymptotically dS case is treated in Sec. III.

\subsection{Conformal transformation of the seed stealth Kerr solution}
As our starting point we consider a subclass of scalar-tensor theories where the speed of gravity equals to the speed of light. 
The most general quadratic DHOST Lagrangian with this property reads~\cite{Langlois:2018dxi},
\begin{equation}
\mathcal{L} = K - G_3 \Box \phi + GR + A_3 \phi^\mu\phi_{\mu\nu}\phi^\nu\Box\phi + A_4 \phi^\mu\phi_{\mu\nu}\phi^{\nu\rho}\phi_\rho + A_5\left(\phi^\mu\phi_{\mu\nu}\phi^\nu\right)^2, \label{eq:lagrangian}
\end{equation}
where $K$, $G_3$, $G$, $A_3$, $A_4$ and $A_5$ are all functions of the scalar field $\phi$ and its kinetic term $X = \phi^\mu\phi_\mu$: $K=K\left(\phi,X\right)$, etc. 
We use the notations $\phi_\mu = \partial_\mu\phi = \nabla_\mu \phi$ and $\phi_{\mu\nu}=\nabla_\mu\nabla_\nu\phi$, while $R$ is the Ricci scalar of the metric $g_{\mu\nu}$. 
The functions $A_4$ and $A_5$ are not free, as they are determined in terms of $G$ and $A_3$,
\begin{equation}
A_4 = \frac{1}{8G}\left(48G_X^2-8\left(G-XG_X\right)A_3-X^2A_3^2\right),\quad A_5 = \frac{A_3}{2G}\left(4G_X+XA_3\right),
\end{equation}
in order to ensure the absence of the Ostrogradski ghost. 
The subscript $X$ implies derivation with respect to $X$. For a start let us consider the shift-symmetric version, where all theory coefficients of (\ref{eq:lagrangian}) depend only on $X$ and not on $\phi$: $K\left(\phi,X\right) = K\left(X\right)$, etc. 

In Ref.~\cite{Charmousis:2019vnf} it has been shown that for shift-symmetric theory~(\ref{eq:lagrangian}) with the following constraints on the functions of the theory,
\begin{equation}
A_3\left(X_0\right) = 0,\quad G_{3X}\left(X_0\right) = 0,\quad \left(K_X+4\Lambda G_X\right)\Bigr\rvert_{X_0} = 0,\quad \left(K+2\Lambda G\right)\Bigr\rvert_{X_0}=0, \label{eq:conditions}
\end{equation}
the field equations admit the Kerr metric (or its dS generalization if $\Lambda\neq 0$) as a solution, provided the scalar field has a constant kinetic term, $\phi^\mu\phi_\mu = X_0$. 
The latter condition amounts to $\phi$ being a Hamilton-Jacobi functional for a geodesic congruence of the spacetime. 
To ensure the regularity of the scalar field throughout the exterior of the black hole, one has to find a geodesic congruence that is defined and is regular everywhere outside the event horizon. 
The stealth solution of Ref.~\cite{Charmousis:2019vnf} is the Kerr metric with a nontrivial scalar field:
\begin{align}
\mathrm{d}s^2={}&-\left(1-\frac{2Mr}{\Sigma}\right)\mathrm{d}t^2+\frac{\Sigma}{\Delta}\mathrm{d}r^2+\Sigma\mathrm{d}\theta^2+\frac{\sin^2\theta}{\Sigma}\Upsilon\mathrm{d}\varphi^2-\frac{4Mar\sin^2\theta}{\Sigma}\mathrm{d}t\mathrm{d}\varphi,\label{eq:kerr}\\
\phi ={}& q\left(t+\int\psi\left(r\right)\mathrm{d}r\right),\label{eq:HJ}
\end{align}
where $M$ the mass, $a$ the angular momentum of the black hole. In the above expression  for the metric the following notations are used,
\begin{equation}
\Sigma=r^2+a^2\cos^2\theta, \quad \Delta = r^2+a^2-2Mr,\quad \Upsilon = \left(r^2+a^2\right)\Sigma+2Mra^2\sin^2\theta, \label{eq:notations}
\end{equation}
while in the expression for the scalar, $q$ is an arbitrary constant with dimension $-1$, and we have defined,
\begin{equation}
\psi\left(r\right)=\frac{\sqrt{2Mr\left(r^2+a^2\right)}}{\Delta}. \label{eq:psi}
\end{equation}
The kinetic term for the scalar field is constant for the above solution, $X = -q^2$, so (\ref{eq:kerr})--(\ref{eq:psi}) is a stealth Kerr solution of the theory (\ref{eq:lagrangian}), provided the conditions (\ref{eq:conditions}) hold, with $X_0=-q^2$ and $\Lambda=0$. 

We now apply a conformal transformation of the form,
\begin{equation}
g_{\mu\nu}\mapsto\widetilde{g}_{\mu\nu}=C\left(\phi\right)g_{\mu\nu}.\label{eq:conformal_map}
\end{equation}
to the metric~(\ref{eq:kerr}). Clearly, the new metric will not be any longer a solution of the theory~(\ref{eq:lagrangian}), instead, the new metric $\widetilde{g}_{\mu\nu}$ with the scalar field~(\ref{eq:HJ}), (\ref{eq:psi}) is a solution of another DHOST theory~\cite{Achour:2016rkg}. 
The new theory has the same form as the original one~(\ref{eq:lagrangian}), with, however, different functions of the theory, see details in Appendix~\ref{app:dhost}. 
We stress that the speed of gravity of the new theory coincides with the speed of light.

The scalar field $\phi$ in our construction plays the role of conformal time, however, it will be more convenient to work with the normalized time $\tau$, defined as 
\begin{equation}
\tau = \frac{\phi}{q}.\label{eq:tau}
\end{equation}
For the sake of convenience, we also introduce a conformal factor in terms of the time $\tau$:
\begin{equation}
C\left(\phi\right)=A^2\left(\phi/q\right)=A^2\left(\tau\right), \label{eq:conformal_scale}
\end{equation}
where the function $A$ will play the role of the asymptotic FLRW conformal scale factor, and it is denoted with an unusual capital letter to avoid confusion with the rotation parameter $a$ of the Kerr metric.
Using~(\ref{eq:HJ}) and (\ref{eq:tau}) it is not difficult to rewrite the metric~(\ref{eq:kerr}) in terms of the new coordinates $\left(\tau,r,\theta,\varphi\right)$, which are the extension of Schwarzschild Painlev\'e-Gullstrand coordinates for the Kerr metric, since by the construction of Ref.~\cite{Charmousis:2019vnf}, $\mathrm{d}\tau$ is the proper time interval of a freely falling particle with vanishing speed at infinity, and vanishing angular momentum\footnote{Note however that only at the absence of rotation, $a=0$, spatial sections of (\ref{eq:conf_kerr}) are flat.}. 
Multiplying the resulting metric by the conformal factor $A^2\left(\tau\right)$ we finally obtain the conformally-related configuration $\left(\widetilde{g}_{\mu\nu},\phi\right)$,
\begin{align}
\mathrm{d}\widetilde{s}^2={}&A^2\left(\tau\right)\left\lbrace -\left(1-\frac{2Mr}{\Sigma}\right)\mathrm{d}\tau^2+\left[\frac{\Sigma}{\Delta}-\psi\left(r\right)^2\left(1-\frac{2Mr}{\Sigma}\right)\right]\mathrm{d}r^2+\Sigma\mathrm{d}\theta^2+\frac{\sin^2\theta}{\Sigma}\Upsilon\mathrm{d}\varphi^2 \right.\nonumber \\ {}&\left.{}-\frac{4Mar\sin^2\theta}{\Sigma}\mathrm{d}\tau\mathrm{d}\varphi+\frac{4Mar\psi\left(r\right)\sin^2\theta}{\Sigma}\mathrm{d}r\mathrm{d}\varphi+2\left(1-\frac{2Mr}{\Sigma}\right)\psi\left(r\right)\mathrm{d}\tau\mathrm{d}r\right\rbrace, \label{eq:conf_kerr}\\
\phi={}& q\tau, \label{eq:phi}
\end{align}
The obtained spacetime~(\ref{eq:conf_kerr}) is axisymmetric but no longer stationary. Expanding the metric as $r\to\infty$ gives
\begin{align}
\mathrm{d}\widetilde{s}^2= A^2\left(\tau\right)\Bigl\{{}&{} -\left[1+\mathcal{O}\left(r^{-1}\right)\right]\mathrm{d}\tau^2 +\left[1+\mathcal{O}\left(r^{-1}\right)\right]\mathrm{d}r^2 + r^2\left[1+\mathcal{O}\left(r^{-2}\right)\right] \mathrm{d}\theta^2+r^2\sin^2\theta\left[1+\mathcal{O}\left(r^{-2}\right)\right] \mathrm{d}\varphi^2 \nonumber\\ {}&{} + \mathcal{O}\left(r^{-1}\right)\mathrm{d}\tau \mathrm{d}\varphi +\mathcal{O}\left(r^{-3/2}\right)\mathrm{d}r \mathrm{d}\varphi+\mathcal{O}\left(r^{-1/2}\right)\mathrm{d}\tau \mathrm{d}r\Bigr\},
\end{align}
which shows that in this limit, the spacetime asymptotes a spatially flat FLRW spacetime written in conformal time $\tau$. 
Note also that for $M=0$, the metric~(\ref{eq:conf_kerr}) reduces exactly to a flat FLRW universe in conformal time $\tau$, although the spatial part is written in ellipsoidal coordinates.
The apparent singularity of the metric~(\ref{eq:conf_kerr}) at $\Delta=0$ is due to a bad behaviour of coordinates there, it can be removed by a change of the coordinate $\varphi$, by defining for instance
\begin{equation}
\mathrm{d}\varphi_+ = \mathrm{d}\varphi + \frac{a}{\Delta}\mathrm{d}r, \label{eq:phi+}
\end{equation}
bringing the metric to the explicitly regular at $\Delta=0$ form,
\begin{equation}
\begin{split}
\mathrm{d}\widetilde{s}^2={}&A^2\left(\tau\right)\left\lbrace -\left(1-\frac{2Mr}{\Sigma}\right)\mathrm{d}\tau^2+\left(2-\frac{1-2Mr/\Sigma}{1+\sqrt{\frac{2Mr}{r^2+a^2}}}\right)\frac{\mathrm{d}r^2}{1+\sqrt{\frac{2Mr}{r^2+a^2}}}+\Sigma\mathrm{d}\theta^2+\frac{\sin^2\theta}{\Sigma}\Upsilon\mathrm{d}\varphi_+^2 \right.
\\ {}&\left.{}-\frac{4Mar\sin^2\theta}{\Sigma}\mathrm{d}\tau\mathrm{d}\varphi_+-2a\sin^2\theta\left(1+\frac{2Mr/\Sigma}{1+\sqrt{\frac{2Mr}{r^2+a^2}}}\right)\mathrm{d}r\mathrm{d}\varphi_++2\left(1-\frac{1-2Mr/\Sigma}{1+\sqrt{\frac{2Mr}{r^2+a^2}}}\right)\mathrm{d}\tau\mathrm{d}r\right\rbrace.\label{eq:good_coord}
\end{split}
\end{equation}
The explicitly regular form of the above metric shows that it is regular everywhere, except the usual Kerr ring singularity at $\Sigma=0$ and possible singularities of the scale factor, e.g. the Big Bang singularity. However, the various square roots in the above expression are not twice differentiable at $r=0$, therefore the spacetime has a curvature singularity at $r=0$, which is a disk singularity comprising the ring singularity $\Sigma=0$. This is discussed in more detail later, see Eq.~(\ref{eq:ric_conf_kerr}), where the Ricci scalar is seen to diverge only at $r=0$ and at $A\left(\tau\right)=0$. The absence of singularity at $\Delta=0$ for the resulting spacetime is not trivial \textit{a priori} and is made possible because, as explained above, $\tau$ is a Painlev\'e-Gullstrand-like time for the Kerr metric. Indeed, in spherical symmetry for instance, it is known~\cite{Mello:2016irl} that the most naive choice for constructing a cosmological conformal Schwarzschild black hole, namely taking the usual coordinates $\left(t,r,\theta,\varphi\right)$ where $g^{rr}=-g_{tt}=1-2M/r$ and multiplying by a scale factor $A^2(t)$, gives a metric with true curvature singularity at the Schwarzschild horizon $r=2M$; while on the other hand, taking Schwarzschild metric in Painlev\'e-Gullstrand coordinates $\left(\tau,r,\theta,\varphi\right)$ and multiplying it by a scale factor $A^2(\tau)$ gives a metric, known as the Culetu spacetime~\cite{Culetu:2012ih}, with only the expected curvature singularities at $r=0$ and $A(\tau)=0$.

In what follows we will illustrate the calculations by choosing a power-law scale factor, 
\begin{equation}
\label{alpha}
A(\tau)=A_0|\tau |^\alpha,
\end{equation}
where it is assumed that $\tau>0$ for $\alpha>0$ and $\tau<0$ for $\alpha<0$. 
In this case our solution mimics asymptotically a Friedmann universe of GR sourced by a perfect fluid with equation of state{\footnote{From the point of view of DHOST theories, $\alpha$ is  a parameter of the theory at hand rather than a parameter of the solution.
As such each $\alpha$ labelling a particular equation of state stands for a different solution of a different DHOST theory. 
} $w=\frac{2-\alpha}{3\alpha}$.
Note that the standard cosmological time $T$ is related to the conformal time via the relation $A(\tau)\mathrm{d}\tau=\mathrm{d}T$. 
Depending on $\alpha$, the latter can be integrated to give
\begin{equation}
\label{eq:T}
T=
\begin{cases}
	\frac{A_0}{|\alpha+1|}|\tau|^{\alpha+1}, & \text{for $\alpha>0$ and $\alpha<-1$} \\
	-A_0\ln\left(-\tau/A_0\right), & \text{for $\alpha=-1$}\\
	T_0-\frac{A_0}{\alpha+1}|\tau|^{\alpha+1}, & \text{for $-1<\alpha<0$},
		 \end{cases}
\end{equation}
where for $\alpha=-1$ the constant $1/A_0$ corresponds to the constant Hubble parameter of de Sitter $H_0=1/A_0$,
and for $-1<\alpha<0$  the constant $T_0$ is the time of the so-called Big Rip~\cite{Caldwell:1999ew}, where the scale-factor of the universe diverges at a finite cosmological time. 
The cosmological time coordinate $T$ extends from $0$ to $+\infty$ with $T=0$ being the Big Bang in the first case in the above expression, in the second case $-\infty<T<+\infty$, and in the last case $T$ goes from $-\infty$ to the Big Rip time $T_0$. 
The scale factor in terms of the cosmological time $T$ is given by
\begin{equation}
\label{eq:AT}
A(T) \propto
\begin{cases}
	T^\frac{\alpha}{\alpha+1}, & \text{for $\alpha>0$ and $\alpha<-1$}\\
	e^{T/A_0}, & \text{for $\alpha=-1$}\\
	\left(T_0-T\right)^\frac{\alpha}{\alpha+1}, & \text{for $-1<\alpha<0$}.
		 \end{cases}
\end{equation}
In particular the choice $\alpha=2,1,-1$ corresponds to matter, radiation and cosmological constant respectively. 

In a nutshell, the metric~(\ref{eq:conf_kerr}) or~(\ref{eq:good_coord}), dressed with the scalar field~(\ref{eq:phi}), is a vacuum solution of a DHOST theory presented in Appendix~\ref{app:dhost}, and is seen to behave as FLRW spacetime when $r\to\infty$. We will now turn on to the characterization of this spacetime as a black hole.

\subsection{Double-null foliations of conformal spacetimes}
\label{sub:folliations}

Since the metric we obtained in~(\ref{eq:conf_kerr}) or (\ref{eq:good_coord}) is neither spherically symmetric nor stationary, understanding its properties is not straightforward. In particular, instead of event horizons, it is more convenient (and more physically relevant) to use the notion of a trapping (apparent) horizon. For the definition of trapping surfaces one needs to introduce {\it{expansion}}, which is related to the behaviour of null geodesic congruences. In order to attack this problem, we will mostly follow the 2+2 formalism initiated by Hayward in Ref.~\cite{Hayward:1993wb}, adapting it to our current purposes when 
needed\footnote{Ref.~\cite{Hayward:1993wb} does not require null congruences to be geodesics. However most works, including the pioneering Ref.~\cite{Penrose:1964wq} or more recent, e.g.~\cite{Faraoni:2015ula} or~\cite{Faraoni:2016xgy} consider only congruences of null geodesics. We adopt the latter, more physical notion.\label{fn:geod}}. 
Note that the usual formalism leads to ambiguities, see~\cite{Faraoni:2016xgy} and references therein, in the identification of trapping horizons. 
In recalling the formalism, we point out sources of ambiguities and propose some prescriptions to get rid of some of them.

In the approach of~\cite{Hayward:1993wb}, the spacetime manifold $\mathcal{M}$ with metric $g_{\mu\nu}$ is foliated by space-like 2-surfaces $\mathcal{S}$ which are the intersection of two families of null 3-surfaces $\Sigma_u$ and $\Sigma_v$. 
The surfaces $\Sigma_u$ and $\Sigma_v$ are defined as the surfaces of constant $u$ and $v$, where $u$ and $v$ are functions on spacetime. 
The normal one-forms $-\mathrm{d}u\equiv L=L_\mu \mathrm{d}x^\mu$ and $-\mathrm{d}v\equiv N=N_\mu \mathrm{d}x^\mu$ are null, and the dual vectors $L^\mu\partial_\mu$ and $N^\mu\partial_\mu$ are future-directed. As explained in footnote~\ref{fn:geod}, we impose these vectors to be geodesic, that is to say, $g^{\mu\nu}L_\mu\nabla_\nu L_\rho\propto L$ and $g^{\mu\nu}N_\mu\nabla_\nu N_\rho\propto N$. In a nutshell, we assume that spacetime has a pair of null coordinates $u$ and $v$, associated to the null geodesic one-forms $L=-\mathrm{d}u$ and $N=-\mathrm{d}v$. Nevertheless, it will turn out that the normalisation of $L$ and $N$ can be crucial for identifying the trapping horizons, see the discussion below after Eq.~(\ref{eq:rescalings}) and in Appendix B. For later purposes, we thus introduce $l=\epsilon^2 L$ and $n=\delta^2 N$ which are just rescalings of $L$ and $N$, with $\epsilon$ and $\delta$ functions on spacetime. The scalar product between two future-directed vectors is negative, thus we write it as
\begin{equation}
g^{\mu\nu}l_\mu n_\nu = -\frac{1}{F^2},\label{eq:scalar_prod}
\end{equation} 
where $F$ is also a function of spacetime.
The first fundamental form is defined as usual,
\begin{equation}
h_{\mu\nu}=g_{\mu\nu}+F^2\left(l_\mu n_\nu+n_\mu l_\nu\right).\label{eq:hmunu}
\end{equation}
The tensor $h_{\mu\nu}$ is the induced metric on $\mathcal{S}$ (the spacelike 2-surfaces orthogonal to both $L$ and $N$); $h_{\mu\nu}$ projects every vector field of $\mathcal{M}$ onto $\mathcal{S}$. The seeked for relevant quantities are then the {\it{expansions}} $\theta_\pm$, defined by
\begin{equation}
\theta_{\pm}=\frac{1}{2}h^{\mu\nu}\mathcal{L}_{\pm}h_{\mu\nu},\label{eq:expansion}
\end{equation}
where $\mathcal{L}_{\pm}$ denotes the Lie derivative with respect to the vector fields $F^2 l^\mu\partial_\mu$ and $F^2 n^\mu\partial_\mu$ respectively. A more explicit form of~(\ref{eq:expansion}) is easily obtained:
\begin{equation}
\theta_+ = F^2\nabla^\mu l_\mu+F^4n^\nu l^\mu\nabla_\mu l_\nu,\quad \theta_- = F^2\nabla^\mu n_\mu+F^4l^\nu n^\mu\nabla_\mu n_\nu.\label{eq:expansions}
\end{equation}
Apart from the expansions~(\ref{eq:expansions}), the other important quantity is the evolution of one expansion along the other geodesic, that is, $\mathcal{L}_-\theta_+$ and $\mathcal{L}_+\theta_-$. 
Indeed one defines a {\it{future outer trapping horizon}} as a 3-surface $H$ on which three properties hold: 
ingoing light rays converge, $\theta_-\rvert_H<0$, while outgoing light rays are parallel on the surface, $\theta_+\rvert_H=0$, and in addition, $\mathcal{L}_-\theta_+\rvert_H<0$, which implies that outgoing light rays are diverging outside and converging inside the surface~\cite{Hayward:1993wb}. 
The existence of a future outer trapping horizon therefore defines a black hole in our cosmological context. More generally, a trapping horizon $H$, defined as a 3-surface on which $\theta_+\rvert_H=0$, is embodied with two properties: future or past; and outer or inner. It is said to be future (past) horizon if $\theta_-\rvert_H<0$ ($\theta_-\rvert_H>0$); 
and it is outer (inner) horizon if $\mathcal{L}_-\theta_+\rvert_H<0$ ($\mathcal{L}_-\theta_+\rvert_H>0$). 
For example, in a Reissner-Nordstr\"om black hole, the outer (event) horizon is a future outer trapping horizon, while the inner (Cauchy) horizon is a future inner trapping horizon. For a maximally-extended Schwarzschild black hole, the black hole horizon is a future outer trapping horizon, while the white hole horizon is a past outer trapping horizon\footnote{These results apply when the normalisations of $l$ and $n$ are chosen appropriately. The discussion which follows and in appendix~B precisely gives counter-examples, which in turn enable to give rules for the proper normalisation.}. In the above definitions, we have fixed the vanishing expansion to be $\theta_+$, but one could obviously rewrite these definitions symmetrically with this time $\theta_-\rvert_H=0$.

The existence and nature of trapping horizons do not depend on the normalization of $l$ and $n$, unless the normalization is singular. Indeed, by rescaling $l\to \gamma^2 l$, $n\to \beta^2 n$  with $\gamma$ and $\beta$ functions on spacetime we have from~(\ref{eq:scalar_prod}) and (\ref{eq:expansions}),
\begin{equation}
F^2\to\frac{F^2}{\gamma^2\beta^2},\quad \theta_+\to\frac{\theta_+}{\beta^2},\quad \theta_-\to \frac{\theta_-}{\gamma^2}\quad \mathcal{L}_-\theta_+\rvert_H\to \frac{\mathcal{L}_-\theta_+}{\gamma^2\beta^2}\Big\rvert_H,\quad \mathcal{L}_+\theta_-\rvert_H\to\frac{\mathcal{L}_+\theta_-}{\gamma^2\beta^2}\Big\rvert_H.\label{eq:rescalings}
\end{equation}
Therefore the trapping horizons are identified unambiguously under such rescaling, unless one of the functions $\gamma$ or $\beta$ diverges or vanishes. 
In Appendix~\ref{app:ln} we give an example of how the singular choice of normalization of $l$ and $n$ can lead to wrong conclusions about the trapping horizons. 

Having in mind the example of Appendix~\ref{app:ln}, as well as the expression~(\ref{eq:expansions}) for the expansions and~(\ref{eq:rescalings}) for their behaviour under rescaling, we therefore give two requirements for the proper normalisation of $l$ and $n$. 
First, we fix their scalar product~(\ref{eq:scalar_prod}) to be finite, in particular in this paper without loss of generality we choose  $F^2=1$. 
This normalization ensures that $F^2$ does not vanish nor diverge, avoiding thus possible unphysical zeros of the expansions according to~(\ref{eq:expansions}). 
Second, we impose $l$ and $n$ to be well-defined in the whole spacetime (apart of course from true curvature singularities). Indeed, a rescaling of the form $l\to\gamma^2 l$, $n\to n/\gamma^2$, although preserving $F^2=1$, could give unphysical vanishing or divergence of the expansions at the roots or poles of $\gamma^2$ according to~(\ref{eq:rescalings}). As a summary, to study the trapping horizons of a spacetime, we first find double null coordinates $u$ and $v$ associated to null geodesic future-directed one-forms $L=-\mathrm{d}u$ and $N=-\mathrm{d}v$, and then define $l$ and $n$, proportional respectively to $L$ and $N$, such that $l$ and $n$ (i) have unit scalar product $g^{\mu\nu}l_\mu n_\nu=-1$ and (ii) are well-defined everywhere (apart from curvature singularities). With this definition, the expansions are then given by
\begin{equation}
\theta_+ = \nabla^\mu l_\mu+n^\nu l^\mu\nabla_\mu l_\nu,\quad \theta_- = \nabla^\mu n_\mu+l^\nu n^\mu\nabla_\mu n_\nu.\label{eq:expansions_bis}
\end{equation}
Using~(\ref{eq:expansions_bis}) along with $\mathcal{L}_-\theta_+ = n^\mu\partial_\mu\theta_+$ and $\mathcal{L}_+\theta_- = l^\mu\partial_\mu\theta_-$, we identify appropriately the future/past, outer/inner trapping horizons $H$ as defined by~\cite{Hayward:1993wb} and recalled above. 
These trapping horizons do not depend on rescaling of $l$ and $n$ preserving conditions (i) and (ii). 
Indeed, the only possible rescaling preserving the scalar product is $l\to\gamma^2 l$, $n\to n/\gamma^2$, and~(\ref{eq:rescalings}) then shows that $\mathcal{L}_\pm\theta_\mp\rvert_H$ are invariant, while $\theta_+\to \gamma^2\theta_+$ and $\theta_-\to \theta_-/\gamma^2$. 
Therefore, given a choice of $L=-\mathrm{d}u$ and $N=-\mathrm{d}v$, the requirements (i) and (ii) enable to identify trapping horizons without ambiguity. The only ambiguity in the identification of trapping horizons of the spacetime thus regards the initial choice of double null geodesic coordinates $u$ and $v$, but we will see that, for the cases of interest in the present article, such an ambiguity does not arise. 

The formalism above is directly applicable to the conformal Kerr metric~(\ref{eq:conf_kerr}), where we have two metrics on the same manifold $\mathcal{M}$ that are conformally related by construction, $\widetilde{g}_{\mu\nu} = A^2g_{\mu\nu}$, and $A$ is a function on spacetime. 
In this case it is easy to relate the 2+2 foliation for $g_{\mu\nu}$ and $\widetilde{g}_{\mu\nu}$ metrics. 
Indeed, the exact one-forms $L=-\mathrm{d}u$ and $N=-\mathrm{d}v$ are defined independently of the metric, while their null norm is preserved by the conformal change.
In addition, the one-forms $L=-\mathrm{d}u$ and $N=-\mathrm{d}v$ are also geodesic in the new metric: since the Christoffel coefficients are modified as
\begin{equation}
\widetilde{\Gamma}^\lambda_{\nu\rho}=\Gamma^\lambda_{\nu\rho}+\frac{\partial_\nu A}{A}\delta^\lambda_\rho+\frac{\partial_\rho A}{A}\delta^\lambda_\nu-\frac{\partial_\sigma A}{A}g^{\sigma\lambda}g_{\nu\rho},
\end{equation} 
one computes
\begin{equation}
\widetilde{g}^{\mu\nu}L_\mu\widetilde{\nabla}_\nu L_\rho = A^{-2}g^{\mu\nu}L_\mu\left\lbrace \nabla_\nu L_\rho -\frac{\partial_\nu A}{A}L_\rho-\frac{\partial_\rho A}{A}L_\nu+\frac{\partial_\sigma A}{A}g^{\sigma\lambda}L_\lambda g_{\nu\rho}\right\rbrace.
\end{equation}
The second and fourth term compensate, while the third vanishes because $L$ is null. If $L$ is geodesic for $g_{\mu\nu}$, i.e. $g^{\mu\nu}L_\mu\nabla_\nu L_\rho\propto L$, one then has $\widetilde{g}^{\mu\nu}L_\mu\widetilde{\nabla}_\nu L_\rho\propto L$. Therefore $L$ is geodesic for $\widetilde{g}_{\mu\nu}$, and the same holds for $N$. Thus we use the same $L$ and $N$ for both spacetimes. Then, if $l\propto L$ and $n\propto N$ are the associated normalized one-forms for $g_{\mu\nu}$, with $g^{\mu\nu}l_\mu n_\nu=-1$, then the one-forms $\widetilde{l}$ and $\widetilde{n}$ for $\widetilde{g}_{\mu\nu}$ are $\widetilde{l}=Al$, $\widetilde{n}=An$ so they are normalized with respect to the metric $\widetilde{g}_{\mu\nu}$, $\widetilde{g}^{\mu\nu}\widetilde{l}_\mu \widetilde{n}_\nu=-1$. If the expansions $\theta_\pm$ for the metric $g_{\mu\nu}$ are known, see~(\ref{eq:expansions_bis}), then the expansions $\widetilde{\theta}_\pm$ for the conformally related metric read,
\begin{equation}
\widetilde{\theta}_+ = \frac{1}{A}\left(\theta_+ + \frac{2}{A}g^{\mu\nu}l_\mu\partial_\nu A\right),\quad \widetilde{\theta}_- = \frac{1}{A}\left(\theta_- + \frac{2}{A}g^{\mu\nu}n_\mu\partial_\nu A\right).\label{eq:thetatilde}
\end{equation}
The case treated in this paper corresponds to $g_{\mu\nu}$ being the stationary, axisymmetric Kerr metric, and $A=A(\tau)$ where $\tau$ is a conformal time. Denoting $\dot{A}\equiv \mathrm{d}A/\mathrm{d}\tau$, the equations $\widetilde{\theta}_\pm\rvert_{\widetilde{H}}=0$, defining trapping horizons $\widetilde{H}$ of the conformal metric $\widetilde{g}_{\mu\nu}$, yield 
\begin{equation}
\frac{\dot{A}}{A}\Big\rvert_{\widetilde{\theta}_+=0} = -\frac{\theta_+}{2l^\tau},\quad \frac{\dot{A}}{A}\Big\rvert_{\widetilde{\theta}_-=0} = -\frac{\theta_-}{2n^\tau},
\end{equation}
where of course $l^\tau=g^{\tau\mu}l_\mu$ and $n^\tau=g^{\tau\mu}n_\mu$. The signs of $\widetilde{\theta}_-$ when $\widetilde{\theta}_+$ vanishes, and vice-versa, determine if the trapping horizons are future or past. They are given by
\begin{equation}
\widetilde{\theta}_-\rvert_{\widetilde{\theta}_+=0}=\frac{1}{A}\left(\theta_--\frac{n^\tau}{l^\tau}\theta_+\right),\quad \widetilde{\theta}_+\rvert_{\widetilde{\theta}_-=0}=\frac{1}{A}\left(\theta_+-\frac{l^\tau}{n^\tau}\theta_-\right).\label{eq:signs_theta}
\end{equation} 
In this paper we will always assume a power-law behaviour of the scale factor $A(\tau)=A_0\left\lvert\tau\right\rvert^\alpha$. Eq.~(\ref{eq:thetatilde}) then implies that the trapping horizons are given by the equations $\tau=\tau_\pm\left(r,\theta\right)$ where
\begin{equation}
\tau_+ = -\alpha\frac{2l^\tau}{\theta_+},\quad \tau_- = -\alpha\frac{2n^\tau}{\theta_-}.\label{eq:tau_plus_minus}
\end{equation}
Finally, for such a scale factor, the Lie derivatives for the conformal spacetime, $\widetilde{\mathcal{L}_\mp\theta_\pm}$, whose signs define whether the trapping horizons are inner or outer, are computed to be
\begin{align}
\widetilde{\mathcal{L}_-\theta_+}\rvert_{\widetilde{\theta}_+=0} ={}&{} \frac{1}{2\alpha l^\tau A^2}\left\lbrace 2\alpha \left(l^\tau n^i\partial_i \theta_+ - \theta_+ n^i\partial_i l^\tau\right)-n^\tau \theta_+^2\right\rbrace,\label{eq:signs_lie_1} \\
\widetilde{\mathcal{L}_+\theta_-}\rvert_{\widetilde{\theta}_-=0} ={}&{} \frac{1}{2\alpha n^\tau A^2}\left\lbrace 2\alpha \left(n^\tau l^i\partial_i \theta_- - \theta_- l^i\partial_i n^\tau\right)-l^\tau \theta_-^2\right\rbrace,\label{eq:signs_lie}
\end{align}
where $l^i = g^{i\mu}l_\mu$ and $n^i = g^{i\mu}n_\mu$ with index $i$ running over spatial coordinates. 
Eqs.~(\ref{eq:signs_theta})--(\ref{eq:signs_lie}) show that, in order to find the trapping horizons and their nature for the conformal spacetime $\widetilde{g}_{\mu\nu} = A^2\left(\tau\right)g_{\mu\nu}$, one needs to know the two expansions $\theta_+$ and $\theta_-$, and the two contravariant vectors $l^\mu$ and $n^\mu$, for the seed metric $g_{\mu\nu}$.

\subsection{Spherically symmetric case: Culetu spacetime}
\label{sub:culetu}

First we consider the simplest case of vanishing angular momentum of a black hole, $a=0$, that is to say, the seed metric~(\ref{eq:kerr}) is the Schwarzschild metric. 
From (\ref{eq:conf_kerr}) and (\ref{eq:phi}) the solution for the conformal metric and the scalar field follows,
\begin{align}
\mathrm{d}\widetilde{s}^2 ={}& A^2(\tau)\Biggl\{-\left(1-\frac{2M}{r}\right)\mathrm{d}\tau^2+2\sqrt{\frac{2M}{r}}\mathrm{d}\tau \mathrm{d}r+\mathrm{d}r^2+r^2\mathrm{d}\Omega^2\Biggr\},\label{eq:sch}\\
\phi ={}& q\tau,\label{eq:HJ_sch}
\end{align}
where $\mathrm{d}\Omega^2$ is the metric of the unit 2-sphere. The metric~(\ref{eq:sch}) was first discussed by Culetu~\cite{Culetu:2012ih}: it is the Schwarzschild metric $\mathrm{d}s^2$ in Painlev\'e-Gullstrand coordinates, multiplied by a scale factor, resulting in $\mathrm{d}\widetilde{s}^2=A^2(\tau)\mathrm{d}s^2$. Note however that in our approach (\ref{eq:sch}) and (\ref{eq:HJ_sch}) is a solution of a DHOST theory, with no matter, i.e. in vacuum. 
The choice of a profile for $A$ corresponds to a conformal factor for the DHOST theory, via Eq.~(\ref{eq:conformal_scale}), i.e. each factor $A=A(\tau)$ (cosmological expansion history) implies a choice of the DHOST theory. 
The Ricci scalar of the Culetu spacetime reads
\begin{equation}
R = \frac{3}{A^2}\left\lbrace 2\frac{\ddot{A}}{A}-\frac{3}{r}\sqrt{\frac{2M}{r}}\frac{\dot{A}}{A}\right\rbrace,
\end{equation}
where $\dot{A}=\mathrm{d}A/\mathrm{d}\tau$, etc. This tells us that the spacetime is singular at $A=0$ (Big Bang singularity) and at $r=0$ ('black hole' singularity). 
If $A(\tau)=A_0|\tau|^\alpha$ ($\alpha>0$), a Big Bang exists at $\tau=0$, and in this case we shall consider the time coordinate to belong to $(0,\infty)$. 
Such a spacetime asymptotically describes a decelerating universe. 
On the contrary when $\alpha<0$ the time coordinate belongs to $(-\infty,0)$, and the spacetime is asymptotically an accelerating homogeneous universe. In particular for $\alpha=-1$ we have de Sitter spacetime in conformal coordinates with $A = -1/\left(H_0\tau\right)$, where the constant $H_0$ is the Hubble rate.  
The Culetu spacetime was recently studied in detail in~\cite{Sato:2022yto} for positive exponents of the scale factor, $\alpha>0$, in the context of General Relativity. 
It was shown that the spacetime solves the Einstein equations only if it is sourced by an energy-momentum tensor violating standard energy conditions. 
This is in contrast to the context of the present paper, in which the Culetu spacetime is a vacuum solution of a scalar-tensor theory for each expansion factor. 
In order to be self-contained, we will re-derive the results for the trapping horizons in the case of Culetu spacetime studied in~\cite{Sato:2022yto} for $\alpha>0$ by using the method described above. In the case of positive $\alpha$ our results fully agree with the findings of~\cite{Sato:2022yto}.

As presented in Sec~\ref{sub:folliations} and Appendix~B, there is a unique (normalized and well-defined) double-null foliation preserving the spherical symmetry of the Schwarzschild spacetime, see Eqs.~(\ref{eq:LN_Sch}) and~(\ref{eq:ln_sch}). 
This applies in particular for Culetu spacetime. 
The expansions $\theta_\pm$ for the Schwarzschild seed spacetime are given by~(\ref{eq:exp_sch}), while the seed contravariant vectors are readily computed to be
\begin{equation}
l^\mu \partial_\mu = \frac{1}{\sqrt{2}}\left[\partial_\tau + \left(1-\sqrt{\frac{2M}{r}}\right)\partial_r\right],\quad n^\mu \partial_\mu = \frac{1}{\sqrt{2}}\left[\partial_\tau - \left(1+\sqrt{\frac{2M}{r}}\right)\partial_r\right].
\end{equation}
Using the above expressions in (\ref{eq:signs_theta}) and (\ref{eq:tau_plus_minus}), one immediately gets that for $A=A_0\left\lvert\tau\right\rvert^\alpha$, the expansions $\widetilde{\theta}_\pm$ vanish at $\tau=\tau_\pm$ with 
\begin{equation}
\tau_+ = -\alpha r\left(1-\sqrt{\frac{2M}{r}}\right)^{-1},\quad \tau_- = \alpha r \left(1+\sqrt{\frac{2M}{r}}\right)^{-1}.\label{eq:taupm}
\end{equation}
Since $\tau$ and $\alpha$ must have the same sign, the trapping horizon $\tau=\tau_+$ must have $r<2M$, while the trapping horizon $\tau=\tau_-$ extends for all $r$. We plot these trapping horizons in Fig.~\ref{fig:1} for positive and negative $\alpha$. 
The graph for $\alpha=-1$ shows that the radial coordinate $r$ for both trapping horizons is shrinking to zero at  $\tau=0$ (future infinity). 
This happens because $r$ is not a physical distance but a comoving coordinate, while the physical radius (as measured by a far away observer) is 
\begin{equation}
R_{phys}=A\left(\tau\right)r. \label{eq:Rphys}
\end{equation}
We thus present in Fig.~\ref{fig:2} the trapping horizons, in terms of the physical radius $R_{phys}$. 
Both physical horizons expand for all presented cases. 
Now, let us examine the nature of these trapping horizons. Using~(\ref{eq:signs_theta}) we find
\begin{figure}
\begin{subfigure}{9cm}
\includegraphics[width=\linewidth]{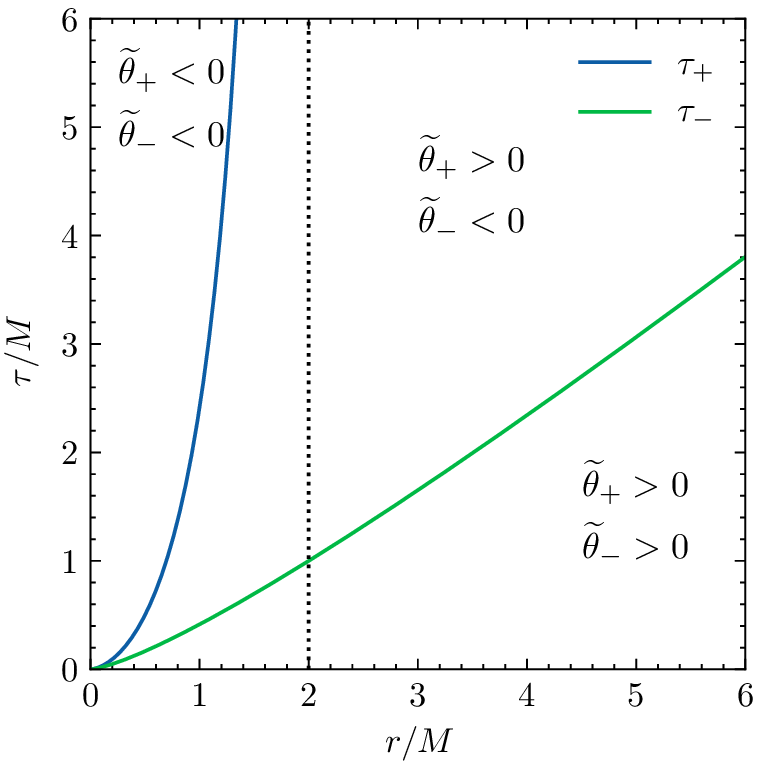}
\end{subfigure}
\begin{subfigure}{8.8cm}
\includegraphics[width=\linewidth]{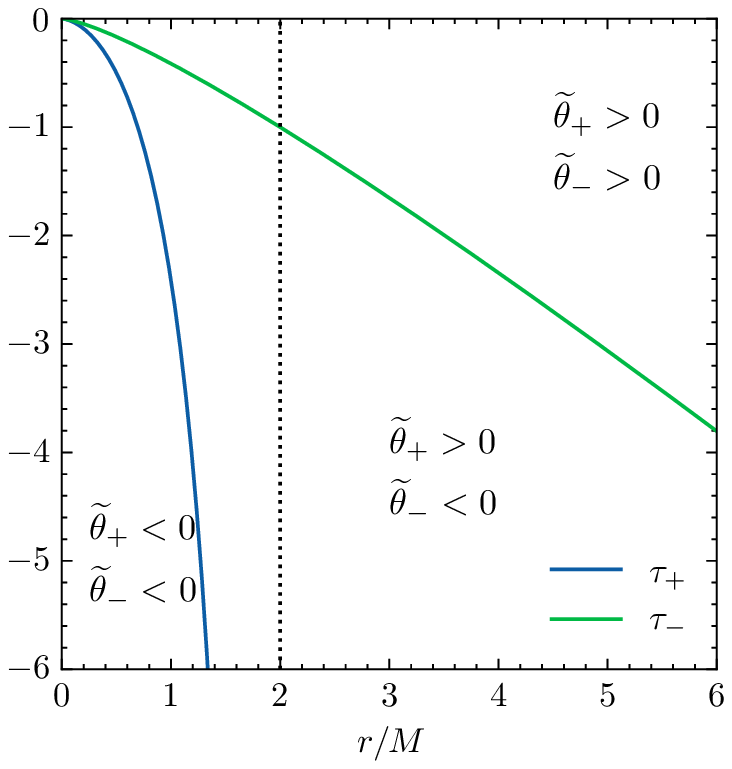}
\end{subfigure}
\caption{Trapping horizons $\tau=\tau_+(r)$ and $\tau=\tau_-(r)$ for the Culetu spacetime, where $r$ is the coordinate radius, for $\alpha=1$ (radiation, left) and $\alpha=-1$ (cosmological constant, right). $\tau_+$ diverges at $r=2M$ which is indicated by the dotted line.}
\label{fig:1}
\end{figure} 
\begin{figure}
\begin{subfigure}{8.9cm}
\includegraphics[width=\linewidth]{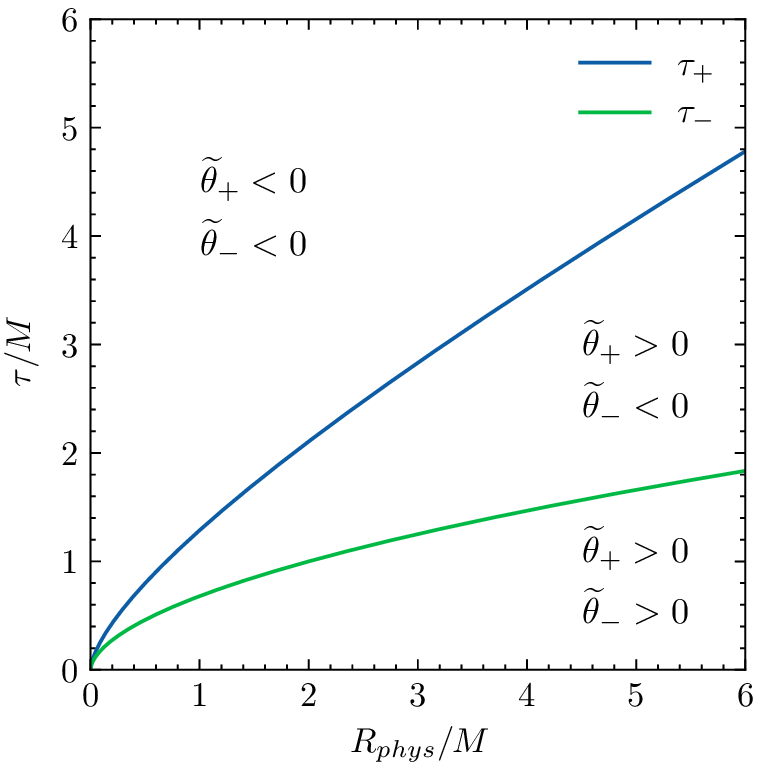}
\end{subfigure}
\begin{subfigure}{8.9cm}
\includegraphics[width=\linewidth]{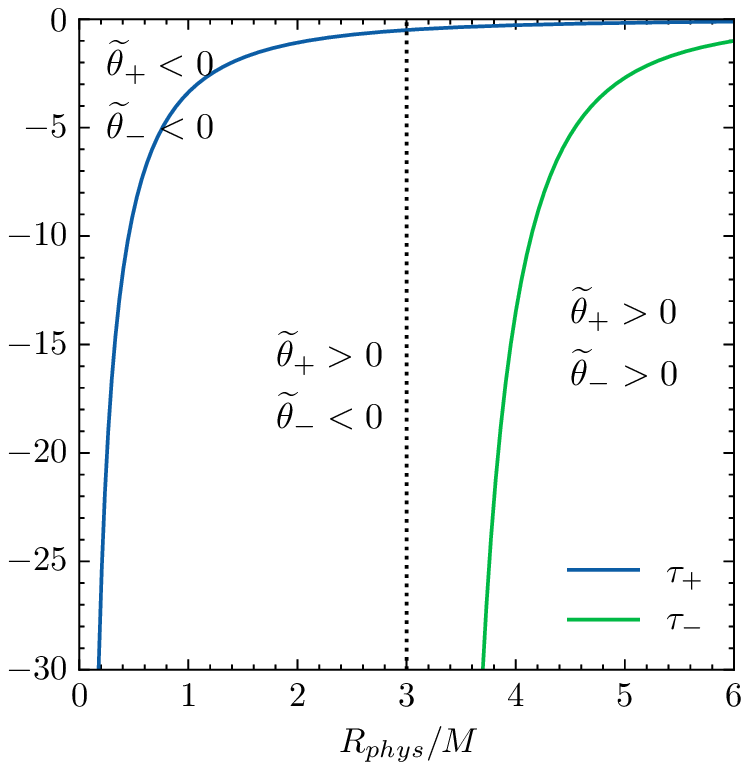}
\end{subfigure}
\caption{Trapping horizons $\tau=\tau_+\left(R_{phys}\right)$ and $\tau=\tau_-\left(R_{phys}\right)$ for the Culetu spacetime, where $R_{phys}=A\left(\tau\right)r$ is the physical radius, for $\alpha=1$ (radiation, left) and $\alpha=-1$ (cosmological constant, right). On this last plot, $\tau_-$ diverges at $R_{phys}=1/H_0$ which is indicated by the dotted line, where $H_0$ is the constant cosmological Hubble rate, taken for the plot to be $H_0=1/\left(3M\right)$. For the left plot, the factor $A_0$ appearing in $A=A_0\tau^\alpha$ is set to unity.}
\label{fig:2}
\end{figure} 
\begin{equation}
\widetilde{\theta}_-\rvert_{\tau_+} = -\frac{2\sqrt{2}}{Ar}<0,\quad \widetilde{\theta}_+\rvert_{\tau_-}= \frac{2\sqrt{2}}{Ar}>0,
\end{equation}
showing that $\tau=\tau_+$ is a future trapping horizon, while $\tau=\tau_-$ is a past trapping horizon. Finally, (\ref{eq:signs_lie_1})--(\ref{eq:signs_lie}) give
\begin{align}
\widetilde{\mathcal{L}_-\theta_+}\rvert_{\tau_+} ={}&{} \frac{1}{\alpha A^2r^2}\left[\alpha\left(1+\sqrt{\frac{2M}{r}}\right)\left(1-\frac{3}{2}\sqrt{\frac{2M}{r}}\right)-\left(1-\sqrt{\frac{2M}{r}}\right)^2\right],\label{eq:lmtp_culetu}\\ 
\widetilde{\mathcal{L}_+\theta_-}\rvert_{\tau_-} ={}&{} \frac{1}{\alpha A^2r^2}\left[\alpha\left(1-\sqrt{\frac{2M}{r}}\right)\left(1+\frac{3}{2}\sqrt{\frac{2M}{r}}\right)-\left(1+\sqrt{\frac{2M}{r}}\right)^2\right].\label{eq:lptm_culetu}
\end{align}
Let us for the moment focus of the case of decelerating universe, $\alpha>0$. 
Since in this case $\tau=\tau_+$ exists for $r<2M$, it is clear from~(\ref{eq:lmtp_culetu}) that $\widetilde{\mathcal{L}_-\theta_+}\rvert_{\tau_+}<0$, therefore $\tau=\tau_+$ is a future outer trapping horizon, and the spacetime is indeed a cosmological black hole. 
As regards the past trapping horizon $\tau=\tau_-$, there are two separate cases depending on whether $\alpha>1$ or $\alpha<1$. It is easy to show from~(\ref{eq:lptm_culetu}) that $\widetilde{\mathcal{L}_+\theta_-}\rvert_{\tau_-}$ has a zero at 
\begin{equation}
r_{1} = \frac{8+\alpha\left(13\alpha-12\right)+\left(4-\alpha\right)\sqrt{\alpha\left(25\alpha-16\right)}}{4\left(\alpha-1\right)^2}M,
\end{equation}
if and only if $\alpha>1$. Therefore, for $0<\alpha\leq 1$, the past trapping horizon is outer, while for $\alpha>1$, it is outer for $r<r_{1}$ and inner for $r>r_{1}$. 

Let us comment on the asymptotic behaviour of the black hole and cosmological horizons at large $\tau$ in the case of decelerating universe, $\alpha>0$. 
From~(\ref{eq:taupm}) and (\ref{eq:Rphys}) one can see that the cosmological (physical) radius behaves at large $\tau$ as 
$$R_c \simeq A(\tau)\frac{\tau}\alpha = \frac{\alpha+1}{\alpha}T,$$ 
where we used Eq.~(\ref{eq:T}) to express $R_c$ in cosmological time $T$. 
This coincides with the expression for the cosmological radius in pure FLRW universe, $R_c=1/H(T)$ where $H(T)=A^{-1}\mathrm{d}A/\mathrm{d}T$, see e.g.~\cite{Faraoni:2015ula}. 
The black hole comoving horizon asymptotically approaches $2M$, implying that the physical radius of the black hole trapping horizon $R_{BH}$ grows proportionally to the scale factor $R_{BH}\simeq 2M A$ at large times.
Note that the cosmological radius expands faster than the black hole horizon.

As for the accelerating universe, $\alpha<0$, for $\alpha<-2/3$ (including asymptotically de Sitter case, $\alpha=-1$), $\tau=\tau_+$ is a future outer trapping horizon (black hole horizon), while the past trapping horizon $\tau=\tau_-$ (cosmological horizon) is outer for $r<r_{2}$ and inner for $r>r_{2}$ where
\begin{equation}
r_{2} = \frac{8+\alpha\left(13\alpha-12\right)-\left(4-\alpha\right)\sqrt{\alpha\left(25\alpha-16\right)}}{4\left(\alpha-1\right)^2}M.
\end{equation}
For $0>\alpha>-2/3$, $\tau=\tau_-$ is a past inner trapping horizon, while the future trapping horizon $\tau=\tau_+$ is inner for $r<r_{2}$ and outer for $r>r_{2}$ (and in this case, $r_{2}<2M$, so there is indeed a region $2M>r>r_{2}$ where $\tau=\tau_+$ is future outer trapping horizon). 
Note that in this case the nature of the future trapping horizon changes with time, being inner at early times and outer later, it is the only range of $\alpha$ for which this happens. 
For $\alpha=-2/3$, $\tau=\tau_+$ is future outer and $\tau=\tau_-$ is past inner. In a word, all values of $\alpha$ lead to the conclusion that the Culetu spacetime possesses a future outer trapping horizon and is therefore a cosmological black hole.

From~(\ref{eq:taupm}) and (\ref{eq:Rphys}) we conclude that the cosmological de Sitter horizon $R_c = H^{-1}_0$ is only recovered in the asymptotic past, while at $\tau\to 0^-$ the cosmological horizon diverges as 
\begin{equation}
R_c \simeq  \frac{1}{H_0}\left[\left(\frac{2M}{-\tau}\right)^{1/3}+\frac{2}{3}\right].
\end{equation}
Such a behaviour can be explained by the influence of the black hole on the cosmology in asymptotic future. 
Indeed, the black hole horizon has the following asymptotic behavior at $\tau\to 0^-$:
\begin{equation}
R_{BH} \simeq  \frac{1}{H_0}\left[\left(\frac{2M}{-\tau}\right)^{1/3}-\frac{2}{3}\right].
\end{equation}
As one can see from the above expressions, the leading behaviour of the two horizon coincide, $R_{BH}\simeq R_c$ at asymptotic future. 
In other words, the black hole expands so fast that it covers almost the whole patch of the universe inside the cosmological radius, thus modifying the behaviour of the cosmological horizon as well.
In general, for the accelerating universe, at $\tau\to 0^-$ the two trapping horizons have the same leading behavior $R_{phys}\propto (-\tau)^{\alpha+2/3}$. 
When $\alpha<-2/3$, the cosmological and black holes horizons diverge at $\tau\to 0^-$, while for $-2/3<\alpha<0$ they shrink to zero. 
The case $\alpha=-2/3$ is marginal, when asymptotically the two horizons tend to a constant value which depends on the parameter $M$.
Note also that for $-1<\alpha <0$ the limit $\tau\to 0^-$ of the conformal time corresponds to a finite cosmological time according to~(\ref{eq:T}) and (\ref{eq:AT}), the finite $T=T_0$ which corresponds to the Big Rip. 
This finding is interesting to compare with the solution found in~\cite{Babichev:2018ubo} for the fate of the black hole in such a universe with our results here. 
In~\cite{Babichev:2018ubo} it was found that a black hole disappears at the moment of the Big Rip. 
In the solution we present here, the black hole also disappears for the range $-2/3<\alpha<0$, while otherwise the radius of the black hole grows up to the moment of the Big Rip or tends to a constant value.

\subsection{Conformal Kerr spacetime}\label{sub:conf_kerr}

We now turn to the general case of conformal Kerr spacetime (\ref{eq:good_coord}) with non-zero rotation. The Ricci scalar for the conformal metric is readily calculated,
\begin{equation}
R = \frac{3}{A^2}\left[2\frac{\ddot{A}}{A}-\frac{3r^2+a^2}{\Sigma\sqrt{r^2+a^2}}\sqrt{\frac{2M}{r}}\frac{\dot{A}}{A}\right].\label{eq:ric_conf_kerr}
\end{equation}
The above expression indicates two singularities of spacetime: a big-bang singularity at $A\left(\tau\right)=0$, and a two-dimensional disk singularity at $r=0$ (which includes the usual ring singularity at $\Sigma=0$ of the Kerr spacetime). 
The disk singularity appears in this case because the seed Kerr metric is multiplied by a scale factor which is a function of the scalar field (\ref{eq:HJ}), and the latter is not twice differentiable at $r=0$, see Eqs.~(\ref{eq:HJ}) and (\ref{eq:psi}). 
We start by identifying 2+2 foliations of Kerr spacetime in Boyer-Lindquist coordinates $(t,r,\theta,\varphi)$, and then write them in coordinates $(\tau,r,\theta,\varphi_+)$. 
As we have seen in~\ref{sub:culetu}, in the spherically-symmetric case there is a unique such foliation if we impose it to respect the spherical symmetry of the spacetime. 
When rotation is present, it is not enough to impose axial symmetry to obtain a unique 2+2 foliation.

Following Sec.~\ref{sub:folliations}, we thus start by looking for an exact, null geodesic one-form $p=p_\mu \mathrm{d}x^\mu=-\mathrm{d}u$. 
As it was shown by Carter in~\cite{Carter:1968rr}, $p$ is given by the following expression\footnote{Strictly speaking, this result holds true if the integral geodesic curves with tangent vector $p^\mu=g^{\mu\nu}p_\nu$ have affine parametrisation, i.e. $p^\mu\nabla_\mu p^\nu=0$. This is always the case up to rescaling of $p$. Note also that the two $\pm$ of Eq.~(\ref{eq:p}) are independent.}
\begin{equation}
p_\mu \mathrm{d}x^\mu = -E \mathrm{d}t+L_z\mathrm{d}\varphi\pm\frac{\sqrt{\mathcal{R}}}{\Delta}\mathrm{d}r\pm\sqrt{\Theta}\,\mathrm{d}\theta, \label{eq:p}
\end{equation}
where
\begin{equation}
\mathcal{R} = \left(E\left(r^2+a^2\right)-aL_z\right)^2-\Delta K,\quad\Theta = K - \sin^2\theta\left(aE-\frac{L_z}{\sin^2\theta}\right)^2,\label{eq:RK}
\end{equation}
with three constants of motion: the energy $E$, the angular momentum $L_z$ and Carter's constant $K$, which guarantees integrability of the geodesic equations.  
Note that $p$ respects the symmetries of the Kerr spacetime, that is, $p_\mu=p_\mu(r,\theta)$. Along each individual geodesic having $p^\mu=g^{\mu\nu}p_\nu$ as tangent vector, $E$, $L_z$ and $K$ are constants. Nevertheless, the geodesic congruence as a whole may a priori have $E$, $L_z$ and $K$ which depend on $r$ and $\theta$ (not however on $t$ or $\varphi$ due to the axial symmetry). 
For example for the well-known 'principal null congruence' of Kerr, see e.g.~\cite{Wald:1984rg}, $E$ is constant throughout spacetime, but $L_z = a E \sin^2\theta$ depends on the $\theta$-angle\footnote{Note that $L_z$ is still a constant along each geodesic of the congruence, since each geodesic lies in a plane of constant $\theta$.}. 

In our construction we require that $p$ is an exact form, $\mathrm{d}p=0$, which implies that $p_t$ and $p_\varphi$ are constants and 
\begin{equation}
\partial_r p_\theta = \partial_\theta p_r. \label{eq:dpegal0}
\end{equation} 
Since $p_t$ and $p_\varphi$ are constants, $E$ and $L_z$ must be constants throughout the spacetime (note the difference with respect to the principal null congruence).
Given the expression of $\Theta$, regularity at the poles $\theta=0,\pi$ then implies $L_z=0$, and we normalize the affine parameter to have $E=1$, without loss of generality. It will be useful for the following to introduce an auxiliary function $k$, related to Carter's ``function'' $K$ as
\begin{equation}
K(r,\theta) = k^2(r,\theta)+a^2\sin^2\theta. \label{eq:Kk}
\end{equation}
Taking into account the conditions above, from~(\ref{eq:p}) and (\ref{eq:RK}) we identify the pair $(L,N)$, with $L$ outgoing and $N$ ingoing, of exact null geodesic congruence in Kerr spacetime\footnote{One could as well have $-k\,\mathrm{d}\theta$ in $L$ and $+k\,\mathrm{d}\theta$ in $N$. This amounts to a change $k\to -k$ coming from the fact that $k$ is defined through its square, Eq.~(\ref{eq:Kk}), and from the fact that $\pm\sqrt{\Theta}=\pm\left\lvert k\right\rvert$. This choice is of course irrelevant.},
\begin{equation}
L = - \mathrm{d}t+\frac{\sqrt{\mathcal{R}}}{\Delta}\mathrm{d}r+k\,\mathrm{d}\theta,\quad N = - \mathrm{d}t-\frac{\sqrt{\mathcal{R}}}{\Delta}\mathrm{d}r-k\,\mathrm{d}\theta,\label{eq:nullpairkerr}
\end{equation}
where $\mathcal{R} = \left(r^2+a^2\right)^2-\Delta \left(k^2+a^2\sin^2\theta\right)$, 
and $k(r,\theta)$ satisfies the following PDE,
\begin{equation}
\sqrt{\mathcal{R}}\partial_r k + k\partial_\theta k = -a^2\sin\theta\cos\theta, \label{eq:pde}
\end{equation}
as a consequence of the condition~(\ref{eq:dpegal0}).
The fact that $L$ and $N$ are exact forms can be explicitly checked~\cite{Arganaraz:2021fpm}, since,
\begin{equation}
L = -\mathrm{d}u,\quad N = -\mathrm{d}v,\quad u=t-r_s,\quad v=t+r_s,\quad r_s\left(r,\theta\right) = \int^r \frac{\sqrt{\mathcal{R}\left(r',\theta=0\right)}}{\Delta\left(r'\right)}\mathrm{d}r'+\int_0^\theta k\left(r,\theta'\right)\mathrm{d}\theta'.
\end{equation}
Eq.~(\ref{eq:pde}) does not define $k=k(r,\theta)$ uniquely, since one needs to supply a boundary condition. 
It admits for example an obvious solution, $k=a\cos\theta$, giving a Carter's function which is constant throughout spacetime, $K=a^2$. This in fact corresponds to the choice made by Hayward~\cite{Hayward:2004ih} (hence the following subscript 'H'), yielding
\begin{equation}
L_{\text{H}} = - \mathrm{d}t+\frac{\sqrt{\mathcal{R}_{\text{H}}}}{\Delta}\mathrm{d}r+a\cos\theta\,\mathrm{d}\theta,\quad N_{\text{H}} = - \mathrm{d}t-\frac{\sqrt{\mathcal{R}_{\text{H}}}}{\Delta}\mathrm{d}r-a\cos\theta\,\mathrm{d}\theta.\label{eq:hay_choice}
\end{equation}
where 
\begin{equation}
\mathcal{R}_{\text{H}} = \left(r^2+a^2\right)^2-\Delta a^2.
\end{equation}
This choice is however singular at the poles as was pointed out recently by Arga\~naraz and Moreschi~\cite{Arganaraz:2021fpm}.
They considered a different choice~\cite{Arganaraz:2021fpm}, with the requirement that the null geodesic be orthogonal to the 2-sphere of radius $r$ when $r\to\infty$. This choice respects the asymptotic spherical symmetry that the Kerr spacetime possesses as $r\to\infty$ and is regular at the poles as we will see.  
We will follow~\cite{Arganaraz:2021fpm} and refer to this choice of coordinates as 'center-of-mass' null coordinates (hence the following subscript 'cm'). 
The 'center-of-mass' null coordinates are found by imposing the following asymptotic boundary condition on $k$,
\begin{equation}
\lim_{r\to\infty}k_{\text{cm}}(r,\theta)=0.\label{eq:arg_choice}
\end{equation}
From~(\ref{eq:pde}) with the boundary condition~(\ref{eq:arg_choice}) it follows that $k$ is vanishing at the poles (and thus the Carter's function $K$ as well),
\begin{equation}
k_{\text{cm}}\left(r,\theta=0\right)=k_{\text{cm}}\left(r,\theta=\pi\right)=0,\quad K_{\text{cm}}\left(r,\theta=0\right)=K_{\text{cm}}\left(r,\pi\right)=0.\label{eq:vanish_poles}
\end{equation}
The function $k_{\text{cm}}$ is anti-symmetric with respect to the equatorial plane, $k_{\text{cm}}(r,\pi-\theta)=-k_{\text{cm}}(r,\theta)$. 
To see this one notices that the function $-k_{\text{cm}}(r,\pi-\theta)$ satisfies the same equation~(\ref{eq:pde}) and the boundary condition~(\ref{eq:arg_choice}), thus the solutions must coincide. 
This immediately leads to an unsurprising symmetry of the Carter's function with respect to the equatorial plane, $K_{\text{cm}}(r,\pi-\theta)=K_{\text{cm}}(r,\theta)$. 
We have thus, in particular, $k_{\text{cm}}(r,\theta=\pi/2)=0$ and $K_{\text{cm}}(r,\theta=\pi/2)=a^2$.

The functions $k_{\text{cm}}$ and $K_{\text{cm}}$ can be found via numerical integration, in order to use the result for the center-of-mass null coordinates for application to  problems in Kerr spacetime, see e.g.~\cite{Arganaraz:2022mks}. 
To be self-contained, we present our numerical integration of~(\ref{eq:pde}) in appendix~\ref{app:cm}, which shows that this integration can be carried out for all $(r,\theta)\in\left[0,+\infty\right)\times\left[0,\pi\right]$, along with some technical but necessary results related to the center-of-mass null coordinates. All inferences we draw from numerical integration have been verified for various angular momenta $a\in\left[0,M\right]$, but for conciseness, all figures are presented for $a=0.5M$. 

Having established that the choice of the boundary condition~(\ref{eq:arg_choice}) indeed leads to a well-defined $k(r,\theta)$, let us return to the pair of general null-forms $\left(L,N\right)$ which satisfy (\ref{eq:nullpairkerr})--(\ref{eq:pde}). For the moment we do not assume $k=k_{\text{cm}}(r,\theta)$, we will use this later. 
In coordinates $\left(\tau,r,\theta,\varphi_+\right)$ of~(\ref{eq:good_coord}), the pair  $\left(L,N\right)$ reads
\begin{equation}
L = -\mathrm{d}\tau +\frac{\sqrt{\mathcal{R}}+\sqrt{2Mr\left(r^2+a^2\right)}}{\Delta}\mathrm{d}r+k\,\mathrm{d}\theta,\quad N = -\mathrm{d}\tau -\frac{\sqrt{\mathcal{R}}-\sqrt{2Mr\left(r^2+a^2\right)}}{\Delta}\mathrm{d}r-k\,\mathrm{d}\theta.
\end{equation}
The associated null pair $(l,n)$, with scalar product equal to $-1$ and well-defined in the whole Kerr spacetime (apart from the curvature singularity) is
\begin{equation}
l = \left(\frac{\Sigma\left(\Sigma-k^2\right)}{2\Upsilon}\right)^{1/2}\frac{L\Delta}{\sqrt{\mathcal{R}}+\sqrt{2Mr\left(r^2+a^2\right)}},\quad n = \left(\frac{\Sigma\left(\Sigma-k^2\right)}{2\Upsilon}\right)^{1/2}\frac{N\Delta}{\sqrt{\mathcal{R}}-\sqrt{2Mr\left(r^2+a^2\right)}}.
\end{equation}
The well-definedness of $l,n$ follows from the fact that $\sqrt{\mathcal{R}}-\sqrt{2Mr\left(r^2+a^2\right)}$ vanishes when $\Delta$ does, see between~(\ref{eq:nullpairkerr}) and~(\ref{eq:pde}). The expansions in the Kerr spacetime are readily computed to be
\begin{equation}
\theta_\pm = \pm\left(\sqrt{\mathcal{R}}\mp \sqrt{2Mr\left(r^2+a^2\right)}\right)F,\label{eq:thetapluskerr}
\end{equation}
where 
\begin{equation}
F = \frac{1}{\sqrt{8\Upsilon\mathcal{R}\Sigma\left(\Sigma-k^2\right)}}\Biggl\{4r\left(r^2+a^2\right)+2\left(M-r\right)K-\Delta\partial_r K+2\sqrt{\mathcal{R}}\left[k\cot\theta+\partial_\theta k\right]\Biggr\}.\label{eq:F}
\end{equation}
For vanishing rotation, $a=0$, $K=0=k$, and one recovers the expansions of Schwarzschild, Eq.~(\ref{eq:exp_sch}). As in the case of Schwarzschild, where expansions~(\ref{eq:exp_sch}) are diverging only at the curvature singularity $r=0$, we expect the expansions of Kerr spacetime to diverge only at the curvature singularity of Kerr, which is at $\Sigma=0$, that is, $r=0$ and $\theta=\pi/2$. Note however that because of the term $k\cot\theta$ in~(\ref{eq:F}) the expansions $\theta_\pm$ diverge at the poles $\theta=0,\pi$ unless if $k$ (and $K$) vanish at the poles. 
In particular, for the null foliations considered  by Hayward, Eq.~(\ref{eq:hay_choice}), the expansions diverge at the poles, since $K_\text{H}=a^2$ as we noted above. 
More generally, as it was underlined in~\cite{Arganaraz:2021fpm}, all previously proposed double null coordinates for Kerr spacetime~\cite{fletcher} suffer from the same problem as the ones by Hayward, due to a conical singularity along the axis of symmetry of the spacelike surfaces $\mathcal{S}$ induced by the 2+2 foliation. 

On the other hand, the expansions $\theta_\pm$ are well-defined at the poles for the center-of-mass foliation of~\cite{Arganaraz:2021fpm}, i.e. when $k=k_\text{cm}$ and $K=K_\text{cm}$, see Eq.~(\ref{eq:vanish_poles}). 
Moreover, in this case, the denominator of $F$ vanishes if and only if $\Sigma=0$ ($F$ has a double pole there), which can be inferred from the properties~(\ref{eq:property_2}-\ref{eq:property_3}) and the definition of $\Upsilon$, Eq.~(\ref{eq:notations}). 
For the center-of-mass foliation, $F$ thus diverges at $r=0$ only for $\theta=\pi/2$, as illustrated by Fig.~\ref{fig:3} in Appendix~\ref{app:plots}, which gathers illustrative plots supporting various results that we affirm in the current section.
This ensures that, for the center-of-mass coordinates, the expansions $\theta_\pm$ are finite in all spacetime but the curvature singularity $\Sigma=0$.
In the rest of the discussion we will therefore focus on the center-of-mass double null coordinates, setting from now $\left(l,n\right)=\left(l_{\text{cm}},n_{\text{cm}}\right)$. 
For brevity, we omit the subscript 'cm' in the following, keeping in mind that we chose the center-of-mass double null coordinates.

The property~(\ref{eq:property}) implies that $\sqrt{\mathcal{R}}-\sqrt{2Mr\left(r^2+a^2\right)}$ in the expression for the expansions~(\ref{eq:thetapluskerr}) has the same sign as $\Delta$ and vanishes if and only if $\Delta$ vanishes\footnote{The expression $\sqrt{\mathcal{R}}-\sqrt{2Mr\left(r^2+a^2\right)}$ also vanishes at the curvature singularity $r=0$ and $\theta=\pi/2$. However $\theta_+$ diverges at the curvature singularity anyway, because $F$ has a double pole there.}. 
On the other hand, numerical integration shows that the factor $F$ of Eq.~(\ref{eq:F}) is positive throughout spacetime, see Fig.~\ref{fig:3} of Appendix~\ref{app:plots}. 
One therefore concludes from Eq.~(\ref{eq:thetapluskerr}) that $\theta_+$ vanishes at the two roots of $\Delta$ (i.e. at the outer and inner horizons of Kerr spacetime, $r_+>r_-$), and it is negative for $r_-<r<r_+$ and positive for $r<r_-$ or $r>r_+$. 
On the other hand  $\theta_-$ is negative through the whole spacetime. 
Therefore this formalism correctly identifies $r=r_+$ and $r=r_-$ as future trapping horizons. 
To see if each of the horizon is outer or inner, we compute the Lie derivative of the expansion,
\begin{equation}
\mathcal{L}_-\theta_+ = -\frac{\sqrt{\mathcal{R}}+\sqrt{2Mr\left(r^2+a^2\right)}}{\sqrt{2\Upsilon\Sigma\left(\Sigma-k^2\right)}}\mathcal{O}\theta_+\label{eq:lmoinsthetaplus}
\end{equation}
where the differential operator $\mathcal{O}$ is
\begin{equation}
\mathcal{O} = \sqrt{\mathcal{R}}\partial_r+k\partial_\theta.\label{eq:op_o}
\end{equation}
$\mathcal{L}_-\theta_+$ is seen from~(\ref{eq:lmoinsthetaplus}) to have opposite sign to $\mathcal{O}\theta_+$. The plot of $\mathcal{O}\theta_+$, Fig.~\ref{fig:3} of Appendix~\ref{app:plots}, shows that $\mathcal{O}\theta_+$ is positive at $r=r_+$ and negative at $r=r_-$. This implies that for Kerr, $r=r_+$ is future outer trapping horizon and $r=r_-$ is future inner trapping horizon, as it should be.

After having established that for the seed Kerr metric the center-of-mass double null foliation yields regular expressions for the expansions, and properly identifies the outer and inner horizons, we turn on to the study of the conformal Kerr spacetime~(\ref{eq:good_coord}). 
We need the seed contravariant vectors,
\begin{align}
l^\mu\partial_\mu ={}&{} \frac{\sqrt{\mathcal{R}}-\sqrt{2Mr\left(r^2+a^2\right)}}{\sqrt{2\Upsilon\Sigma\left(\Sigma-k^2\right)}}\left\lbrace \frac{G_+}{\Delta}\partial_\tau+\mathcal{O}\right\rbrace+l^{\varphi_+}\partial_{\varphi_+},\\
n^\mu\partial_\mu ={}&{} \frac{\sqrt{\mathcal{R}}+\sqrt{2Mr\left(r^2+a^2\right)}}{\sqrt{2\Upsilon\Sigma\left(\Sigma-k^2\right)}}\left\lbrace \frac{G_-}{\Delta}\partial_\tau-\mathcal{O}\right\rbrace+n^{\varphi_+}\partial_{\varphi_+},
\end{align}
where, for brevity, we used the expression $\mathcal{O}$ of Eq.~(\ref{eq:op_o}), and we have not explicitly written the components along $\varphi_+$ since they do not play any role in the following calculations. In the above expressions we have also introduced
\begin{equation}
G_\pm = \Delta\Sigma\pm\sqrt{2Mr\left(r^2+a^2\right)}\left[\sqrt{\mathcal{R}}\pm\sqrt{2Mr\left(r^2+a^2\right)}\right].\label{eq:gpm}
\end{equation}
It is not difficult to prove that the ratio $G_-/\Delta$ is well-defined and positive, and numerical analysis, see Fig.~\ref{fig:3} of Appendix~\ref{app:plots}, shows that $G_+$ is positive. 
By the use of Eq.~(\ref{eq:tau_plus_minus}), assuming $A=A_0\left\lvert\tau\right\rvert^\alpha$, we find that the expansions $\widetilde{\theta}_\pm$ of the conformal Kerr spacetime vanish at $\tau=\tau_\pm$ with
\begin{equation}
\tau_\pm = \frac{-\alpha\sqrt{2}}{\sqrt{\Upsilon\Sigma\left(\Sigma-k^2\right)}}\frac{\left[\sqrt{\mathcal{R}}\mp \sqrt{2Mr\left(r^2+a^2\right)}\right]G_\pm}{\theta_\pm\Delta}. 
\label{eq:tau_conf_kerr}
\end{equation}
From the above expression we infer, that since $\theta_-$ is negative through the whole spacetime, and $\tau$ and $\alpha$ have the same sign, the trapping horizon $\tau=\tau_-\left(r,\theta\right)$ extends for all $r$ and $\theta$. 
On the other hand, as we found above, $\theta_+$ is negative only for $r_-<r<r_+$. 
As a consequence, the trapping horizon $\tau=\tau_+\left(r,\theta\right)$ must be located between $r_-$ and $r_+$. 
Fig~\ref{fig:4} shows $\tau_+$ and $\tau_-$ in terms of the comoving radius $r$ for $\alpha=1$. 
The profiles of $\tau_\pm$ for other values of $\alpha$ differ from the one plotted in Fig.~\ref{fig:4} by a factor $\alpha$, as it is clear from~(\ref{eq:tau_conf_kerr}).
As we underlined above, for the seed Kerr metric $\theta_\pm$ diverge only when $\Sigma=0$, which manifests through Eq.~(\ref{eq:tau_conf_kerr}) in such a way that $\tau_+$ does not vanish, while $\tau_-$ vanishes if and only if $\Sigma=0$. 
Note here the difference with the spherically-symmetric case: both Culetu and conformal Kerr spacetimes have a curvature singularity at $r=0$ (see Eq.~(\ref{eq:ric_conf_kerr})), however for Culetu metric $\tau_+$ and $\tau_-$ vanish at $r=0$, while for conformal Kerr metric $\tau_+$ does not vanish and $\tau_-$ vanishes only at $r=0$ and $\theta=\pi/2$. 
This property of the conformal Kerr metric has a crucial impact on behavior of the curve $\tau_\pm$ close to $R_{phys}=0$, as demonstrated in Fig~\ref{fig:5}. 
For positive $\alpha$, both trapping horizons start to exist a finite amount of time after the Big Bang $\tau=0$, except the past trapping (cosmological) horizon  $\tau=\tau_-$ at $\theta=\pi/2$, which exists at all times. Both horizons expand infinitely $R_{phys}\to +\infty$ as $\tau\to+\infty$. 
In the case of $\alpha=-1$, the future trapping (black hole) horizon $\tau=\tau_+$ has an upper bound and it ceases to exist at late times for all $\theta$, while the past (cosmological) horizon collapses to zero size at late times, unless $\theta=\pi/2$ which still extends up to $R_{phys}\to +\infty$. Let us determine the nature of these trapping horizons, that we have already anticipated in the last few sentences. Eq.~(\ref{eq:signs_theta}) yields
\begin{figure}[b]
\includegraphics[width=0.6\textwidth]{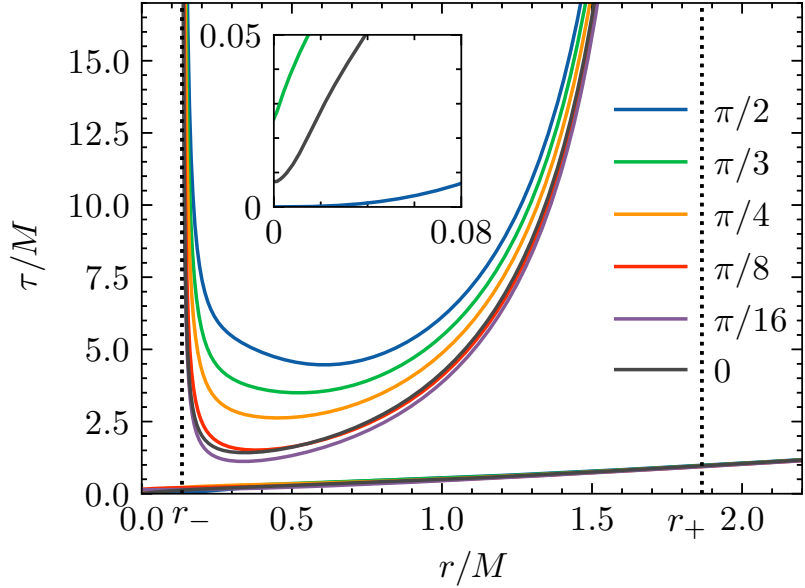}
\caption{Trapping horizons $\tau=\tau_+\left(r,\theta\right)$ and $\tau=\tau_-\left(r,\theta\right)$ for different fixed values of angle $\theta$ indicated by the colors, for the conformal Kerr spacetime with $a=0.5M$, where $r$ is the coordinate radius. $\tau_+$ (curves with a minimum) diverges at $r=r_+$ and $r=r_-$ which are indicated by the dotted lines. $\tau_-$ (cluster of bottom curves) vanishes only at $r=0$ and $\theta=\pi/2$, as is exemplified by the zoom on the curves $\theta=\pi/2$, $\pi/3$ and $0$ near $\tau_-=0$. The plot is obtained for an exponent $\alpha=1$ in the scale factor.}
\label{fig:4}
\end{figure} 
\begin{figure}
\begin{subfigure}{9.1cm}
\includegraphics[width=\linewidth]{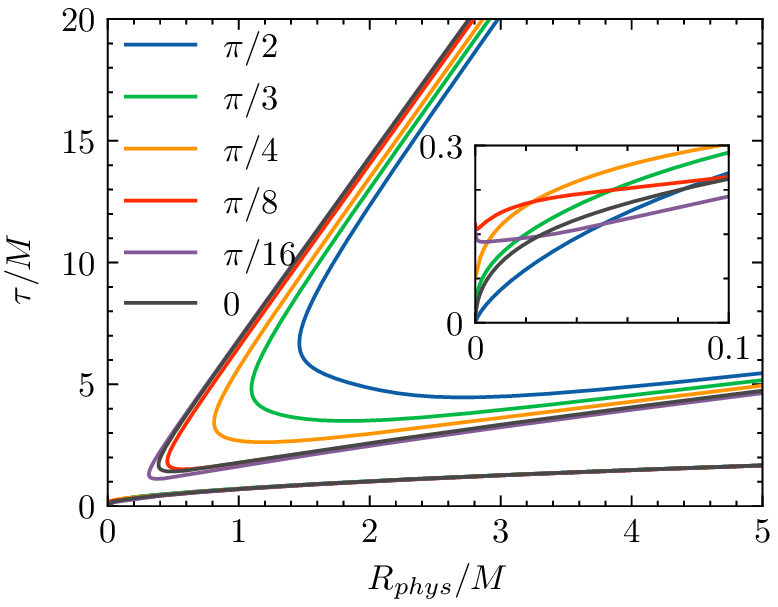}
\end{subfigure}
\begin{subfigure}{8.7cm}
\includegraphics[width=\linewidth]{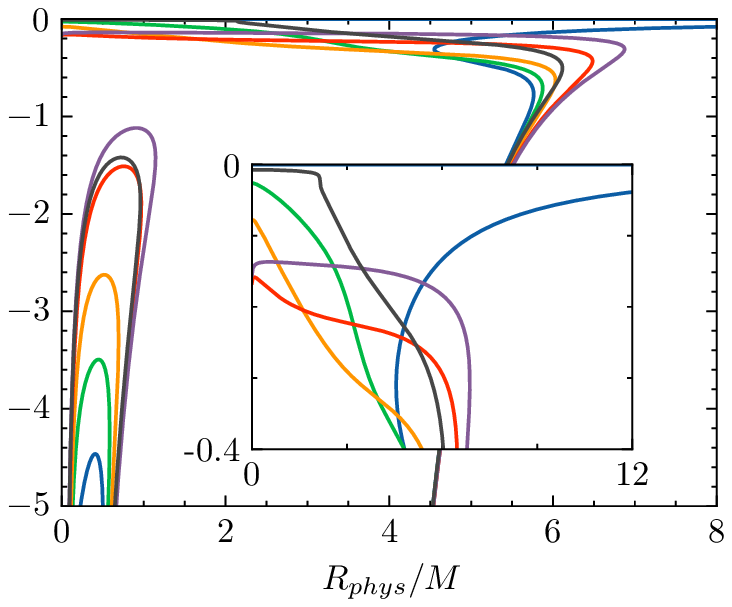}
\end{subfigure}
\caption{Trapping horizons $\tau=\tau_+\left(R_{phys},\theta\right)$ and $\tau=\tau_-\left(R_{phys},\theta\right)$ for different values of angle $\theta$, indicated by the colors (same for both panels), for the conformal Kerr spacetime with $a=0.5M$, where $R_{phys}=A\left(\tau\right)r$ is the physical radius, for $\alpha=1$ (radiation, left plot) and $\alpha=-1$ (cosmological constant, right plot). On the left, $\tau_-$ corresponds to the bottom curves and is zoomed on near $\tau_-=0$, and the factor $A_0$ appearing in $A=A_0\tau^\alpha$ is set to unity. On the right, $\tau_-$ corresponds to the right curves and is zoomed on near $\tau_-=0$, and the Hubble rate is set to $H_0=1/\left(3M\right)$.}
\label{fig:5}
\end{figure} 
\begin{equation}
\widetilde{\theta}_-\rvert_{\tau_+} =\frac{\theta_-}{A}\left( 1+\frac{G_-}{G_+}\right),\quad \widetilde{\theta}_+\rvert_{\tau_-}= \frac{\theta_+}{A}\left( 1+\frac{G_+}{G_-}\right).
\end{equation}
The ratio $G_-/G_+$ has the same sign as $\Delta$ and one can prove that $\left\lvert G_-/G_+\right\rvert<1$. Taking into account the signs of $\theta_\pm$, we conclude that $\widetilde{\theta}_-\rvert_{\tau_+}<0$ and $\widetilde{\theta}_+\rvert_{\tau_-}>0$, that is, $\tau=\tau_+$ is a future trapping horizon and $\tau=\tau_-$ is a past trapping horizon, for any $\alpha$.

Finally, $\widetilde{\mathcal{L}_-\theta_+}\rvert_{\tau_+}$ and $\widetilde{\mathcal{L}_+\theta_-}\rvert_{\tau_-}$ can be computed from~(\ref{eq:signs_lie_1})--(\ref{eq:signs_lie}), however the explicit expression is not very instructive. 
Let us focus on the future trapping horizon $\tau=\tau_+$, and on the three interesting cases of cosmology, $\alpha=2,1,-1$. 
In spherical symmetry, these cases led to $\widetilde{\mathcal{L}_-\theta_+}\rvert_{\tau_+}<0$, that is, the future trapping horizon was outer. Here, since $r=r_+$ and $r=r_-$ are respectively outer and inner for the seed Kerr metric, and given the form of $\tau_+$, see Fig.~\ref{fig:4}, we rather expect to find an outer part close to $r_+$, and an inner part close to $r_-$. 
The plots of Fig.~\ref{fig:6} in Appendix~\ref{app:plots} confirm that this is indeed the case: there exists a radius $r_m(\theta)$ such that for $r_m<r<r_+$, $\widetilde{\mathcal{L}_-\theta_+}\rvert_{\tau_+}<0$, while for $r_-<r<r_m$, $\widetilde{\mathcal{L}_-\theta_+}\rvert_{\tau_+}>0$, that is, the future trapping horizon $\tau=\tau_+$ is outer for $r_m<r<r_+$ and inner for $r_-<r<r_m$. 

 Before moving on to the next section, let us briefly comment on the influence of the rotation parameter $a$. The qualitative picture presented above is the same for any $0<a<M$. When $a$ increases towards $M$, the curve $\tau=\tau_+$ of Fig.~\ref{fig:4} moves upwards, and also the range of $r$ for which it exists decrease, because $r_-$ and $r_+$ get closer. The consequence on the plots of Fig.~\ref{fig:5} is that, for positive $\alpha$, the black hole horizon starts to exist at later times when $a$ increases; while for $\alpha=-1$, the black hole horizon disappears earlier when $a$ increases. If $a\geq M$, the expansion $\theta_+$ for the seed Kerr spacetime, see~(\ref{eq:thetapluskerr}), is positive, therefore, the time $\tau_+$ defined by Eq.~(\ref{eq:tau_conf_kerr}) always has opposite sign to $\alpha$ and there is no black hole horizon. 

\section{Cosmological Kerr-de Sitter black holes}
We have focused in Sec.~II on conformal Kerr spacetimes, starting from the seed stealth Kerr solution in DHOST theory.
In this section we consider a seed stealth Kerr-de Sitter solution, i.e. the seed metric having de Sitter asymptotics, rather than flat ones. 
In~\cite{Charmousis:2019vnf} it was shown that it is possible to construct such a solution with a regular scalar field at both future black hole and cosmological event horizons. 
The Kerr-dS seed stealth solution has the following concise form,
\begin{align}
\mathrm{d}s^2 ={}&{} -\frac{\Delta_r}{\Xi^2\Sigma}\left[\mathrm{d}t-a\sin^2\theta \mathrm{d}\varphi\right]^2+\Sigma\left(\frac{\mathrm{d}r^2}{\Delta_r}+\frac{\mathrm{d}\theta^2}{\Delta_\theta}\right)+\frac{\Delta_\theta\sin^2\theta}{\Xi^2\Sigma}\left[a \mathrm{d}t-\left(r^2+a^2\right)\mathrm{d}\varphi\right]^2,\label{eq:kerrds}\\
\phi ={}&{} q\left\lbrace \frac{\eta}{\Xi}t-\sign\left(r-r_0\right)\int_{r_0}^r\frac{\sqrt{\left[r^2+a^2\right]\left[\eta^2\left(r^2+a^2\right)-\Delta_r\right]}}{\Delta_r}\mathrm{d}r \,\pm a\int\frac{\sin\theta\sqrt{\Delta_\theta-\eta^2}}{\Delta_\theta}\mathrm{d}\theta\right\rbrace, \label{eq:phids}
\end{align} 
where the function '$\sign$' has value $1$ or $-1$ according to the sign of its argument, and
\begin{equation}
\Delta_r = \left(1-H^2r^2\right)\left(r^2+a^2\right)-2Mr,\quad \Delta_\theta= 1+a^2H^2\cos^2\theta,\quad \Xi=1+a^2H^2,\quad H=\sqrt{\frac{\Lambda}{3}},
\end{equation}
while $\eta\in(0,1]$ is a number defined such that $\eta^2\left(r^2+a^2\right)-\Delta_r$ has a double root $r_0$ between the black hole horizon and the cosmological horizon. 
This choice of $\eta$, along with the function '$\sign$', ensure regularity of the scalar field, see details in~\cite{Charmousis:2019vnf}. 
The scalar field $\phi$ behaves like a retarded time at the black hole horizon and like an advanced time at the cosmological horizon, with finite corrections depending on $r$ (see below). 
Hence $\phi$ is regular at both future horizons. 
Eqs.~(\ref{eq:kerrds})--(\ref{eq:phids}) is a solution to the theory~(\ref{eq:lagrangian}), provided conditions~(\ref{eq:conditions}) hold, with $X_0=-q^2$ and $\Lambda=3H^2$. 

Following the same procedure as in Sec.~II, we obtain a solution to the conformally-related DHOST theory presented in appendix~A, for the metric $\mathrm{d}\widetilde{s}^2$ which is conformal to $\mathrm{d}s^2$ and the conformal factor depends on $\phi$. 
In contrast to the case of the Kerr seed metric, for Kerr-dS seed solution, $\phi$ does not correspond exactly to (conformal) cosmological time at large distances. 
The explicit expressions are cumbersome and not particularly enlightening for the general case. 
Therefore let us  concentrate on the case of vanishing rotation, which captures all the issues related to the construction of the conformal metric in this case, being much simpler to analyse. 
We first have a closer look at the scalar field and compare the time defined as $T=\phi/q$ with the cosmological time at spatial asymptotic infinity. 
Eq.~(\ref{eq:phids}) reads in the case $a=0$,
\begin{equation}
T=\eta t-\sign\left(r - r_0 \right) \int_{r_0}^r \frac{\sqrt{\eta^2-f(r)}}{f(r)}\mathrm{d}r,  \label{eq:tau_sds}
\end{equation}
where $f=1-2M/r-H^2r^2$ and, in this case, $\eta$ and $r_0$ have explicit expressions,
\begin{equation}
\eta = \sqrt{1-3\left(MH\right)^{2/3}},\quad r_0 = \left(\frac{M}{H^2}\right)^{1/3}.\label{eq:def_eta}
\end{equation}
One can also define in a standard way a time $T_c$, which is regular at the future de Sitter horizon and approximates the cosmological time at spatial infinity, 
\begin{equation}
{T_c} = t -\int \frac{\sqrt{1-f(r)}}{f(r)}\mathrm{d}r.
\end{equation}
Expressing $T$ in terms of $T_c$ we find that
\begin{equation}
\label{eq:tautauc}
T = \eta\left[T_c + \int_{r_0}^r\frac{\mathrm{d}r}f\left( \sqrt{1-f(r)}-\sign(r-r_0)\sqrt{1-\frac{f(r)}{\eta^2}} \right)  \right].
\end{equation}
First, one can check from the above expressions, that at the cosmological horizon $r_c$, such that $f(r_c)=0$, we have $T=\eta T_c+\text{const}$, implying that $T$ is regular at $r_c$ along with $T_c$.
At spatial asymptotic infinity one finds from~(\ref{eq:tautauc}),
\begin{equation}
T \approx \eta T_c +(1-\eta)\frac{\log(Hr)}{H}, \quad \text{as $r\to\infty$}.
\label{eq:Tcosm}
\end{equation}
We will see in a moment that the logarithmic in $r$ difference of the two times have consequences on the form of the metric. 
Eq.~(\ref{eq:kerrds}) for $a=0$, in terms of the time $T$ reads,
\begin{equation}
\mathrm{d}s^2=-\frac{f\left(r\right)}{\eta^2}\mathrm{d}T^2 - 2\sign(r-r_0)\frac{\sqrt{\eta^2-f(r)}}{\eta^2}\mathrm{d}T\mathrm{d}r+\frac{\mathrm{d}r^2}{\eta^2}+r^2\mathrm{d}\Omega^2 \label{eq:conf_sds}.
\end{equation}
The consequent change of the radial coordinate $r=\eta\rho\exp\left(HT\right)$ brings the seed metric to the following form,
\begin{equation}
\mathrm{d}s^2=  -\left(f+H^2r^2-2H^2r^2W\right)\frac{\mathrm{d}T^2}{\eta^2}+ \frac{2Hr}{\eta}\mathrm{e}^{HT}W \mathrm{d}T \mathrm{d}\rho+\mathrm{e}^{2HT}\left(\mathrm{d}\rho^2+\eta^2\rho^2 \mathrm{d}\Omega^2\right),\label{eq:kdstau}
\end{equation}
where $r$ must be understood in terms of $\rho$ and $T$, and
\begin{equation}
\label{eq:w}
W = 1-\frac{\sign(r-r_0)}{H r}\sqrt{\eta^2-f(r)}\Bigg|_{r=\eta\rho\exp\left(HT\right)}
\end{equation}
In order to obtain the conformal metric one simply multiplies~(\ref{eq:kdstau}) by $A^2(T)$. 
However, the seed metric~(\ref{eq:kdstau}) is not yet in a form that allows us to interpret the conformal metric easily, because $T$ plays a role of a cosmic time,  it can be seen from~(\ref{eq:kdstau}). 
On the other hand an extra factor $A^2(T)$ that we multiply by the seed metric is conformal. 
Therefore we change $T$ to conformal time,
\begin{equation}
T = - \frac{1}H \log\left(-H\tau\right).\label{eq:Ttau}
\end{equation}
Finally, multiplying~(\ref{eq:kdstau}) by the conformal factor, we obtain the following scalar-tensor conformal Schwarzschild-de Sitter solution to the DHOST theory presented in appendix A with conformal factor $C\left(\phi\right)=A^2\left(\phi/q\right)$:
\begin{align}
\mathrm{d}\widetilde{s}^2={}&{}\frac{A^2\left(T\right)}{H^2\tau^2}\left\{ -\left(f(r)+H^2r^2-2H^2r^2W(r)\right)\frac{\mathrm{d}\tau^2}{\eta^2} 
	+ \frac{2Hr}{\eta}W(r) \mathrm{d}\tau \mathrm{d}\rho+\mathrm{d}\rho^2+\eta^2\rho^2 \mathrm{d}\Omega^2\right\},\label{eq:fullds}
\\
\phi ={}&{}\frac{q}{H}\log\left(-\frac{1}{H\tau}\right),\label{eq:fullphi}
\end{align}
where it is understood that $T$ is expressed in terms of $\tau$ using~(\ref{eq:Ttau}) and $r=-\eta\rho /(H\tau)$, while $W(r)$ is given by~(\ref{eq:w}).
One can see an extra factor $\eta^2$ in front of the 2-sphere of the metric, which introduces a subtlety in an interpretation of the metric at spacial asymptotic infinity.
Indeed from the above expression we obtain the asymptotic expression for the conformal metric as $\rho\to \infty$,
\begin{equation}
\mathrm{d}\widetilde{s}^2\approx \frac{A^2\left(T\right)}{H^2\tau^2}\left( -\mathrm{d}\tau^2 + \mathrm{d}\rho^2+\eta^2\rho^2\mathrm{d}\Omega^2\right),\label{eq:almost_conformal}
\end{equation}
which does not coincide with the usual expression for FLRW metric in conformal coordinates, due to the factor $\eta^2$ in the last term.
There seem to be therefore a solid angle deficit. 
The reason for appearing $\eta$ is the difference of the time defined by $\phi$ and the standard cosmological time, we have observed before, see Eq.~(\ref{eq:Tcosm}).
In fact this 'solid deficit' can be removed by change of coordinates to the new coordinates $\left\{\tau_c,\rho_c\right\}$, $\tau=\tau_c \left( H\rho_c \right)^{\eta-1}$, $\rho =H^{\eta-1}\rho_c^\eta/\eta$. Note that the new time $\tau_c$ is defined analogously to $\tau$, see~(\ref{eq:Ttau}): $T_c = -H^{-1} \log\left(-H\tau_c\right)$. 
In these coordinates the asymptotic metric~(\ref{eq:almost_conformal}) reads,
\begin{equation}
\mathrm{d}\widetilde{s}^2\approx \frac{A^2\left(T\right)}{H^2\tau_c^2}\left( -\mathrm{d}\tau_c^2 + \mathrm{d}\rho_c^2+\rho_c^2\mathrm{d}\Omega^2\right). \label{eq:FRLW_conformal}
\end{equation}
where $T$ is a function of $\tau_c$ and $\rho_c$: 
\begin{equation}
T=-\frac{1}{H}\log\left( -H\tau_c \right) +\frac{1-\eta}{H}\log\left( H\rho_c \right).
\end{equation}
The above considerations demonstrate that in fact the presence of the solid angle deficit in~(\ref{eq:fullds}) does not influence asymptotic behaviour of the metric at very large distances, i.e. asymptotically at $\rho\to \infty$ the FLRW solution is recovered. 
One can see this differently, e.g. the Ricci scalar of the full conformal metric (\ref{eq:fullds}) taken at $\rho\to \infty$ reads,
\begin{equation}
R=\frac{12 H^2}{A^2}\left( 1+\frac{\tau\left( 4\dot{A}+\tau\ddot{A} \right)}{2A} \right)  
-\frac{3\left( 1-\eta^2 \right) H^2 \tau^3 \dot{A}}{\eta^2 A^3 \rho^2}+ \mathcal{O}(\rho^{-4}),\label{eq:Ricci}
\end{equation}
where the dots denote derivative with respect to $\tau$: $\dot{A}= \frac{\mathrm{d}A(T(\tau))}{\mathrm{d}\tau}$, $\ddot{A}= \frac{d^2A(T(\tau))}{\mathrm{d}\tau^2}$.
The first term in~(\ref{eq:Ricci}) does not depend on $\eta$, it coincides with the expression for the Ricci invariant in the case of homogeneous FLRW (it corresponds to setting $M\to0$ and $\eta\to 1$ in~(\ref{eq:fullds})). The second term is the first inhomogeneous correction to the homogeneous universe, however this correction dies out as $\rho\to \infty$. 
The decay of the inhomogeneity due to a non-zero $(1-\eta^2)$ is however rather weak: as a comparison, the first non-zero correction to~(\ref{eq:Ricci}) containing $M$ (apart from its implicit appearance in $\eta$) decays much faster, with $\mathcal{O}(\rho^{-5})$ order.

A simple example of the general expression~(\ref{eq:fullds}) is when $A=\text{const}$.  
The solution of the conformal metric can be interpreted in this case as a Schwarzschild-de Sitter solution with the rescaling $H\to A H$, $r\to r/A$, $M\to M/A$.
The existence of such a solution can be also seen directly from the new action after the redefinition by the conformal constant factor.

\section{Conclusions}
In this article, we have constructed and analysed the properties of rotating black holes embedded in an FLRW background. 
The main ingredients in this construction are twofold: first, there exists a stealth Kerr solution in scalar-tensor (DHOST) theory~\cite{Charmousis:2019vnf}, with a timelike scalar field; second, if $\left(g_{\mu\nu},\phi\right)$ is solution of a DHOST theory, then $\left(Cg_{\mu\nu},\phi\right)$ is solution of another DHOST theory~\cite{Achour:2016rkg}, where the conformal factor $C$ depends in general on $\phi$ and its kinetic term. 
We applied the conformal transformation to the seed stealth Kerr metric with a conformal factor that explicitly depends on the scalar field\footnote{Note that including the scalar field kinetic term in $C$ would not change anything since this kinetic term is constant for the stealth Kerr metric.}, $C=C\left(\phi\right)$. 
Being timelike, the scalar $\phi$ defines a new time coordinate $\tau\propto\phi$, therefore the conformal factor $C(\phi)$ plays the role of an FLRW scale factor $A\left(\tau\right)$ squared  at asymptotic spatial infinity. This method yields a solution~(\ref{eq:conf_kerr})--(\ref{eq:phi}) to a new DHOST theory, which is explicitly given in Appendix~\ref{app:dhost}. 
Notably the new theory has speed of gravitational waves equal to the speed of light, just as the seed DHOST theory. 
The new metric has the form $\widetilde{g}_{\mu\nu}=A^2(\tau)g^\text{Kerr}_{\mu\nu}$, where  $g^\text{Kerr}_{\mu\nu}$ is the Kerr metric written using $\tau$, Eq.~(\ref{eq:conf_kerr}).
For the Kerr metric we have argued that the time $\tau$ is an analogue of a Painlev\'e-Gullstrand time. 
Crucially, thanks to regularity of $\tau$ at the Kerr horizons, the obtained metric $A^2(\tau)g^\text{Kerr}_{\mu\nu}$ is also regular there, with curvature singularities present only at $A(\tau)=0$ and at $r=0$.

In our construction the conformal metric $\widetilde{g}_{\mu\nu}$ asymptotes FLRW at large distances, with the scale factor  $A(\tau)$ and $\tau$  being the conformal time. 
This happens naturally since at large distances the seed metric is asymptotically flat, and by construction we multiply the seed metric by a conformal factor which depends on $\tau$.
It should be stressed that each choice of the conformal factor corresponds to a different DHOST theory. 
In particular, each $\alpha$ used to parametrize the scale factor in~(\ref{alpha}) implies a different theory which we have specified in Appendix~A.

While we easily recover the asymptotic FLRW at spatial infinity, it is not as obvious to show that the obtained solution is a black hole.
For the non-stationary metric under consideration, the relevant notion to consider is the trapping horizon, with the help of which a black hole is defined by the existence of a future outer trapping horizon. 
This notion is recalled in Sec.~\ref{sub:folliations}, based on the formalism of double null foliations of spacetime~\cite{Hayward:1993wb}, \cite{dInverno:1980kaa}.
Using specific conditions for the null foliations, we then establish how to unambiguously determine the location and nature (e.g. future or past, outer or inner) of the trapping horizons for a conformally-related spacetime using only quantities of the seed spacetime and the conformal factor.

As a warm-up, we first considered vanishing rotation. 
The conformal Kerr metric $\widetilde{g}_{\mu\nu}$ reduces in that case to the spherically-symmetric Culetu spacetime, which is simply conformal to Schwarzschild in Painlev\'e-Gullstrand coordinates. 
This spacetime, along with other conformal spacetimes with different slicings of the Schwarzschild metric, was recently studied as a cosmological black hole in~\cite{Sato:2022yto}, for conformal scale factors $A(\tau)=A_0\left\lvert\tau\right\rvert^\alpha$ with positive $\alpha$, i.e. decelerating universe. 
In~\cite{Sato:2022yto} this spacetime was treated from the perspective of GR with exotic energy-momentum tensor which is calculated from the Einstein equations. 
In this paper we obtain the Culetu spacetime as a vacuum solution of a DHOST theory. 
Our findings for trapping horizons coincide with~\cite{Sato:2022yto} for decelerating universe, with both cosmological and black hole horizons expanding infinitely at late times.
In addition, we studied the case of accelerating universe, i.e. negative $\alpha$. 
In this case, we found that both horizons either expand infinitely or shrink to zero size at late times depending on the value of $\alpha$. 
In particular, for asymptotically de Sitter universe ($\alpha=-1$), both horizons expand to infinity as the conformal time approaches $0^-$. 
Notably, as it was underlined in~\cite{Sato:2022yto}, the study of this spherically-symmetric Culetu spacetime appears straightforward since there exists a unique double null foliation of spacetime respecting the spherical symmetry.

For non-vanishing rotation, the conformal Kerr metric is not spherically symmetric, however, we showed that the relevant double null foliations can be cast in a simple form~(\ref{eq:nullpairkerr}) which depends on a unique function $k(r,\theta)$ solving the PDE~(\ref{eq:pde}). 
The freedom in the choice of the foliation stems from the freedom of choosing boundary conditions for this PDE. 
In fact, in a recent important work~\cite{Arganaraz:2021fpm}, Arga\~naraz and Moreschi proposed a boundary condition on $k(r,\theta)$ which not only is natural, 
since it ensures that the foliation is asymptotically spherically symmetric following the symmetries of Kerr metric, 
but in addition, up to date, is the only condition resulting in an everywhere smooth double null foliation of the Kerr spacetime. 
We therefore used the double null foliation of Arga\~naraz and Moreschi, dubbed 'center-of-mass double null coordinates', to study the trapping horizons of the conformal Kerr spacetime. 
The non-vanishing rotation introduces features which are not present in the case of spherical symmetry: in particular, for decelerating universe, the trapping horizons start to exist a finite time after the Big Bang, while for accelerating universe, they shrink to zero size or cease to exist at late times\footnote{Apart from the equatorial plane, where the behaviour is similar to the spherically-symmetric case.}, see Sec.~\ref{sub:conf_kerr}. 

To sum up, we demonstrated that the conformal transformation of the seed stealth Kerr solution of~\cite{Charmousis:2019vnf} indeed leads to rotating black holes embedded in FLRW universe, which are vacuum solutions of DHOST theory. 
We also extended our analysis to the case of a Kerr-de-Sitter seed metric to construct a solution $\widetilde{g}_{\mu\nu}=C(\phi)g^\text{Kerr-dS}_{\mu\nu}$ to a scalar-tensor theory. 
The main difference is that the scalar field is not directly related to a conformal or cosmological time as before, Eq.~(\ref{eq:fullds}) . 
Another interesting feature is that the metric~(\ref{eq:fullds}) contains a factor $\eta^2$ in front of the 2-sphere of the metric, usually associated with a solid deficit angle.
However, asymptotically at large distances the spacetime still behaves as an FLRW universe, with a global conformal scale factor which is the product of two scale factors: the scale factor $A^2=C$ of the conformal transformation, and the scale factor of the seed metric.

The present work opens interesting perspectives for future research. 
It sets a general technique to construct and study the properties of cosmological black holes as vacuum solutions of scalar-tensor theories, by starting with a seed solution dressed with a timelike scalar field. Although in the present paper we used a stealth metric as a seed solution, it is not necessary to start with a stealth spherically symmetric or stationary solution. 
An interesting direction could be the use of a non-stealth spherically-symmetric solution dressed with a time-dependent scalar, for example such as given in~\cite{Charmousis:2021npl}, with possibly different properties for the conformal metric than the one presented here. This also holds for non stealth stationary metrics such as the one presented in \cite{Anson:2020trg}. 
In addition, it would be interesting to study thermodynamics of the apparent horizons of the exact cosmological black hole solutions given in the paper.

\acknowledgments
We would like to thank Mokhtar Hassaine for useful discussions. This work was supported by the French National Research Agency via Grant No. ANR-20-CE47-0001 associated with the project COSQUA (Cosmology and Quantum Simulation). EB and NL acknowledge support of ANR grant
StronG (ANR-22-CE31-0015-01). The work of NL is supported by the
doctoral program Contrat Doctoral Sp\'ecifique Normalien \'Ecole
Normale Sup\'erieure de Lyon (CDSN ENS Lyon).

\appendix 
\section{DHOST theory for the conformally-related metric}
\label{app:dhost}
Starting with the DHOST Lagrangian $\mathcal{L}$ of Eq.~(\ref{eq:lagrangian}), and the associated action $S\left[g_{\mu\nu},\phi\right]=\int d^4x\sqrt{-g}\mathcal{L}$, we introduce a conformally-related metric,
\begin{equation}
g_{\mu\nu}\mapsto\widetilde{g}_{\mu\nu}=C\left(\phi\right)g_{\mu\nu}.
\end{equation}
In this article we introduce the scale factor $A$ and the definition $C\left(\phi\right)=A^2\left(\phi/q\right)$. One then has the following equality,
\begin{equation}
S\left[g_{\mu\nu},\phi\right] = \widetilde{S}\left[\widetilde{g}_{\mu\nu},\phi\right],
\end{equation}
where the new action $\widetilde{S}\left[\widetilde{g}_{\mu\nu},\phi\right]$ belongs to the same class (\ref{eq:lagrangian}) of DHOST theories with $c_g=c$, that is to say,
\begin{equation}
\widetilde{S}\left[\widetilde{g}_{\mu\nu},\phi\right] = \int\mathrm{d}^4x\sqrt{-\widetilde{g}}\widetilde{\mathcal{L}}, \quad \widetilde{\mathcal{L}} = \widetilde{K} - \widetilde{G_3} \widetilde{\Box \phi} + \widetilde{G}\widetilde{R} + \widetilde{A_3} \widetilde{\phi^\mu}\widetilde{\phi_{\mu\nu}}\widetilde{\phi^\nu}\widetilde{\Box\phi} + \widetilde{A_4} \widetilde{\phi^\mu}\widetilde{\phi_{\mu\nu}}\widetilde{\phi^{\nu\rho}}\widetilde{\phi_\rho} + \widetilde{A_5}\left(\widetilde{\phi^\mu}\widetilde{\phi_{\mu\nu}}\widetilde{\phi^\nu}\right)^2.
\end{equation}
In this equation, it is understood that indices are contracted with the metric $\widetilde{g}_{\mu\nu}$, and $\widetilde{\phi_{\mu\nu}} = \widetilde{\nabla}_\mu\widetilde{\nabla}_\nu\phi$, etc. 
The coefficients of the new theory, being the functions of $\phi$ and of the new kinetic term $\widetilde{X} = \widetilde{\phi^\mu}\widetilde{\phi_\mu}$, read,
\begin{align}
\widetilde{K} ={}& \frac{K}{C^2}+\frac{\widetilde{X}C_\phi}{C^2}G_3 + \frac{3\widetilde{X}}{C^2}\left(C_{\phi\phi}-\frac{3C_\phi^2}{2C}\right)G - \frac{\widetilde{X}^3C_\phi^2}{2C}A_3+\frac{\widetilde{X}^3C_\phi^2}{4C}A_4 + \frac{\widetilde{X}^4C_\phi^2}{4}A_5-\widetilde{X}\left(H_3+H_4+H_5\right)_\phi,\\
\widetilde{G_3} ={}& \frac{G_3}{C}-\frac{3C_\phi}{C^2}G-\frac{\widetilde{X}^2C_\phi}{2}A_3 + H_3+H_4+H_5,\\
\widetilde{G} ={}& \frac{G}{C},\\
\widetilde{A_3} ={}& CA_3,\\
\widetilde{A_4} ={}& CA_4,\\
\widetilde{A_5} ={}& C^2A_5,
\end{align}
where the subscript $\phi$ means derivation with respect to $\phi$, and the three following functions are introduced for convenience,
\begin{equation}
H_3 = -\frac{C_\phi}{2}\int\mathrm{d}\widetilde{X} \widetilde{X}A_3,\quad H_4 = \frac{C_\phi}{2}\int\mathrm{d}\widetilde{X} \widetilde{X}A_4,\quad H_5 = \frac{CC_\phi}{2}\int\mathrm{d}\widetilde{X} \widetilde{X}^2A_5.
\end{equation}
Note that these formulas are true for any initial theory (\ref{eq:lagrangian}), without assuming shift-symmetry. 
The conformal mapping~(\ref{eq:conformal_map}) is invertible provided that $C\left(\phi\right)$ has finite non-zero values. 
On the other hand, when $C\left(\phi\right)$ is zero or diverging, the conformal spacetime is singular. 
Therefore, if $\left(g_{\mu\nu},\phi\right)$ is a regular solution of the field equations of $S$ in a patch of the spacetime, then $\left(\widetilde{g}_{\mu\nu},\phi\right)$ is also a regular solution of the field equations of $\widetilde{S}$ in the same patch of the spacetime.
As a consequence, the cosmological black holes we construct in the article are solutions of the equations following from the variational principle of a DHOST theory with $c_g=c$ in vacuum.  Note that no matter Lagrangian is added to the DHOST Lagrangian.

\section{Normalization of null vectors $l$ and $n$ in spherical symmetry}
\label{app:ln}

In this appendix, we discuss conditions on the normalization of $l$ and $n$ which prevent ambiguities in the identification of trapping horizons, with the illustrative example of Schwarzschild spacetime in Painlev\'e-Gullstrand coordinates,
\begin{equation}
\mathrm{d}s^2 = -\left(1-\frac{2M}{r}\right)\mathrm{d}\tau^2+2\sqrt{\frac{2M}{r}}\mathrm{d}\tau \mathrm{d}r+\mathrm{d}r^2+r^2\left(\mathrm{d}\theta^2+\sin^2\theta \mathrm{d}\varphi^2\right).
\end{equation}
The spacetime has spherical symmetry, and it is therefore natural to look for a double-null foliation which respects this symmetry, that is to say, with $L=-\mathrm{d}u$ and $N=-\mathrm{d}v$ orthogonal to the coordinate vectors $\partial_\theta$ and $\partial_\varphi$. Up to global rescalings, a unique such pair exists,
\begin{equation}
L_\mu \mathrm{d}x^\mu = -\mathrm{d}\tau+\frac{\mathrm{d}r}{1-\sqrt{2M/r}},\quad N_\mu \mathrm{d}x^\mu = -\mathrm{d}\tau-\frac{\mathrm{d}r}{1+\sqrt{2M/r}},\label{eq:LN_Sch}
\end{equation}
with associated null coordinates
\begin{equation}
u = \tau - \left(r+2\sqrt{2Mr}+4M\log\left\lvert\sqrt{\frac{r}{2M}}-1\right\rvert\right),\quad v = \tau + r-2\sqrt{2Mr}+4M\log\left\lvert\sqrt{\frac{r}{2M}}+1\right\rvert.
\end{equation}
Applying the formalism of Sec.~\ref{sub:folliations} without rescaling $L$ and $N$, that is, taking $l=L$ and $n=N$, one readily computes $F^2 = \left(r-2M\right)/\left(2r\right)$ and $\theta_+ = \left(r-2M\right)/r^2$ and $\theta_- = -\theta_+$. Both the expansions vanish at $r=2M$, while one expects an expansion to vanish and the other to be negative. In view of~(\ref{eq:rescalings}), one can even choose worse normalisations, like $l=\left(r-2M\right)L$ and $n=\left(r-2M\right)N$, leading to $F^2=1/\left(2r\left(r-2M\right)\right)$ and $\theta_+=1/r^2=-\theta_-$ which do not vanish at all. One might argue that in both these cases, $F^2$ either diverges or vanishes at the horizon, which might be the source of the problems. However, problematic cases can arise even if $F^2$ is regular, say $F^2=1$, by setting for example $l = \left(r-2M\right)^2/\left(2r\right)L$ and $n=N/\left(r-2M\right)$. Then $\theta_+=\left(r-2M\right)^2/r^2$ and $\theta_-=-2/\left(r\left(r-2M\right)\right)$: $\theta_+$ vanishes appropriately at $r=2M$, but $\theta_-$ diverges towards $+\infty$ if $r=\left(2M\right)^-$ and towards $-\infty$ if $r=\left(2M\right)^+$. In this latest case, although $F^2$ and $l$ are regular, $n$ is diverging at $r=2M$. If, finally, one requires that $F^2=1$, and $l$ and $n$ are both regular (apart of course at the spacetime singularity $r=0$), the Schwarzschild horizon is correctly identified as a future outer trapping horizon. If indeed
\begin{equation}
l=\frac{1}{\sqrt{2}}\left(1-\sqrt{\frac{2M}{r}}\right)L,\quad n = \frac{1}{\sqrt{2}}\left(1+\sqrt{\frac{2M}{r}}\right)N,\label{eq:ln_sch}
\end{equation}  
then $F^2=1$, both $l$ and $n$ are well-defined if $r\neq 0$, and one gets
\begin{equation}
\theta_\pm = \pm\frac{\sqrt{2}}{r}\left(1\mp\sqrt{\frac{2M}{r}}\right),\label{eq:exp_sch}
\end{equation}
giving $\theta_-<0$ everywhere, so in particular at $r=2M$ which is the unique vanishing point of $\theta_+$, and in addition one computes $\mathcal{L}_-\theta_+=-\left(2M\right)^{-2}<0$ at $r=2M$, hence a future outer trapping horizon at $r=2M$.

\section{Carter's function for the center-of-mass null coordinates}
\label{app:cm}
This appendix gathers analysis of the Carter's function $K_\text{cm}(r,\theta)$ and of the auxiliary function $k_\text{cm}(r,\theta)$ needed for constructing null geodesic foliations of Sec.~\ref{sub:conf_kerr}.
The auxiliary function $k_\text{cm}(r,\theta)$ is defined via the relation $K_\text{cm} = k_\text{cm}^2+a^2\sin^2\theta$ and it solves the PDE with the boundary conditions corresponding to the center-of-mass coordinates~\cite{Arganaraz:2021fpm},
\begin{equation}
\sqrt{\mathcal{R}_\text{cm}}\partial_r k_\text{cm} + k_\text{cm}\partial_\theta k_\text{cm} = -a^2\sin\theta\cos\theta,\quad \lim_{r\to\infty}k_\text{cm}(r,\theta)=0,\label{eq:pde_appendix}
\end{equation}
where $\mathcal{R}_\text{cm} = \left(r^2+a^2\right)^2-\Delta \left(k_\text{cm}^2+a^2\sin^2\theta\right)$. All plots are presented for an angular momentum $a=0.5M$. 
To numerically solve the above PDE we introduce the radial coordinate $\xi=1/r$, following~\cite{Arganaraz:2021fpm}. 
The boundary condition at $r\to\infty$ is translated to a boundary condition at $\xi=0$. 
Then we solve~(\ref{eq:pde_appendix}) using a Runge-Kutta 4 (RK4) step along $\xi$. 
The drawback of this method is the difficulty to reach $r=0$, which corresponds to $\xi\to\infty$. 
To avoid this problem, one may try to solve along $\theta$ by specifying correctly boundary conditions at $\theta=0$. 
Indeed, the original boundary condition implies the vanishing of $k_\text{cm}$ at the poles, see Eq.~(\ref{eq:vanish_poles}), therefore one might want to impose numerically $k_\text{cm}\left(r,\theta=0\right)=0$ and then make the RK4 step along $\theta$. Unfortunately, the initial (i.e. at $\theta=0$) RK4 step along $\theta$ is then ill-defined because isolating $\partial_\theta k_\text{cm}$ in~(\ref{eq:pde_appendix}) requires to divide by $k_\text{cm}$, which vanishes at $\theta=0$. 

The best strategy therefore seems to make the RK4 step along $\xi$, from $\xi=0$ to very large value $\xi=\xi_\infty$, such that the corresponding value of $r_0=1/\xi_\infty$ be less than the precision of the solver, typically $r_0\lesssim 10^{-16}$. 
Then within the accuracy of the method we assume that $k_\text{cm}\left(r_0,\theta\right)\approx k_\text{cm}\left(r=0,\theta\right)$. 
To verify the consistency of this approximation, we take this numerical $k_\text{cm}\left(r_0,\theta\right)\approx k_\text{cm}\left(r=0,\theta\right)$ as the initial condition for an RK4 solving of~(\ref{eq:pde_appendix}), from $r=0$ to large $r$. 
At finite values of $r$, both solutions, the latter one obtained with RK4 along $r$ and the former one obtained with RK4 along $\xi$, coincide very well. 
For example at the outer horizon $r=r_+$, the typical relative error is of the order of a percent, if we exclude the point $\theta=\pi/2$ where the relative error gets bigger because of the vanishing of $k$, but where the absolute error remains small, of the order of $10^{-4}$. 

Our numerical results suggest that the solution $k_\text{cm}(r,\theta)$ of~(\ref{eq:pde_appendix}) exists for all $(r,\theta)\in\left[0,+\infty\right)\times\left[0,\pi\right]$. 
Fig~\ref{fig:7} shows the numerical solutions $k_\text{cm}(r,\theta)$ and $K_\text{cm}(r,\theta)$ as functions of both $r$ and $\theta$. 
Fig.~\ref{fig:8} displays the behaviour of $k_\text{cm}$ for fixed $r$ and varying $\theta$ and vice-versa.
\begin{figure}
\begin{subfigure}{8.9cm}
\includegraphics[width=\linewidth]{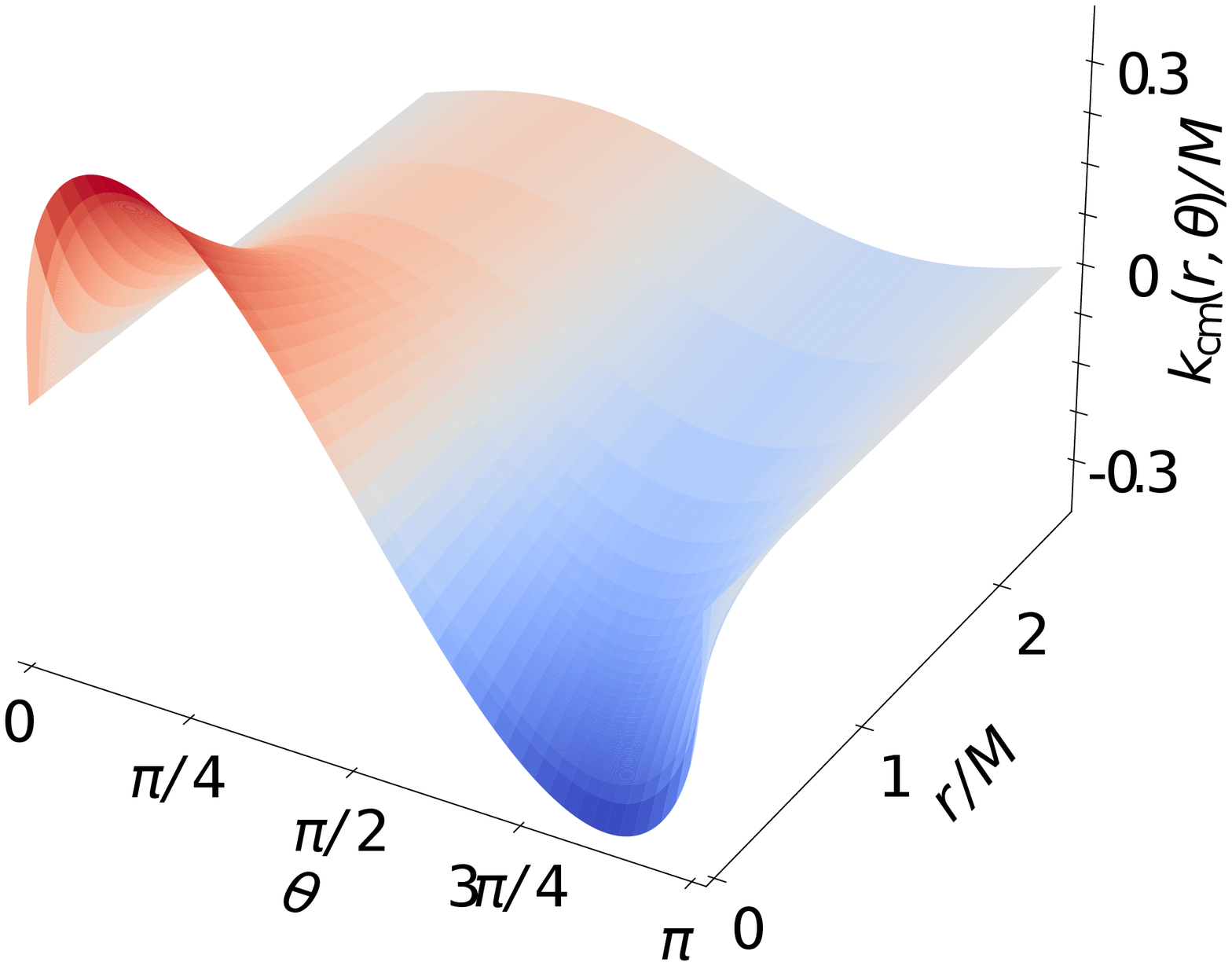}
\end{subfigure}
\begin{subfigure}{8.9cm}
\includegraphics[width=\linewidth]{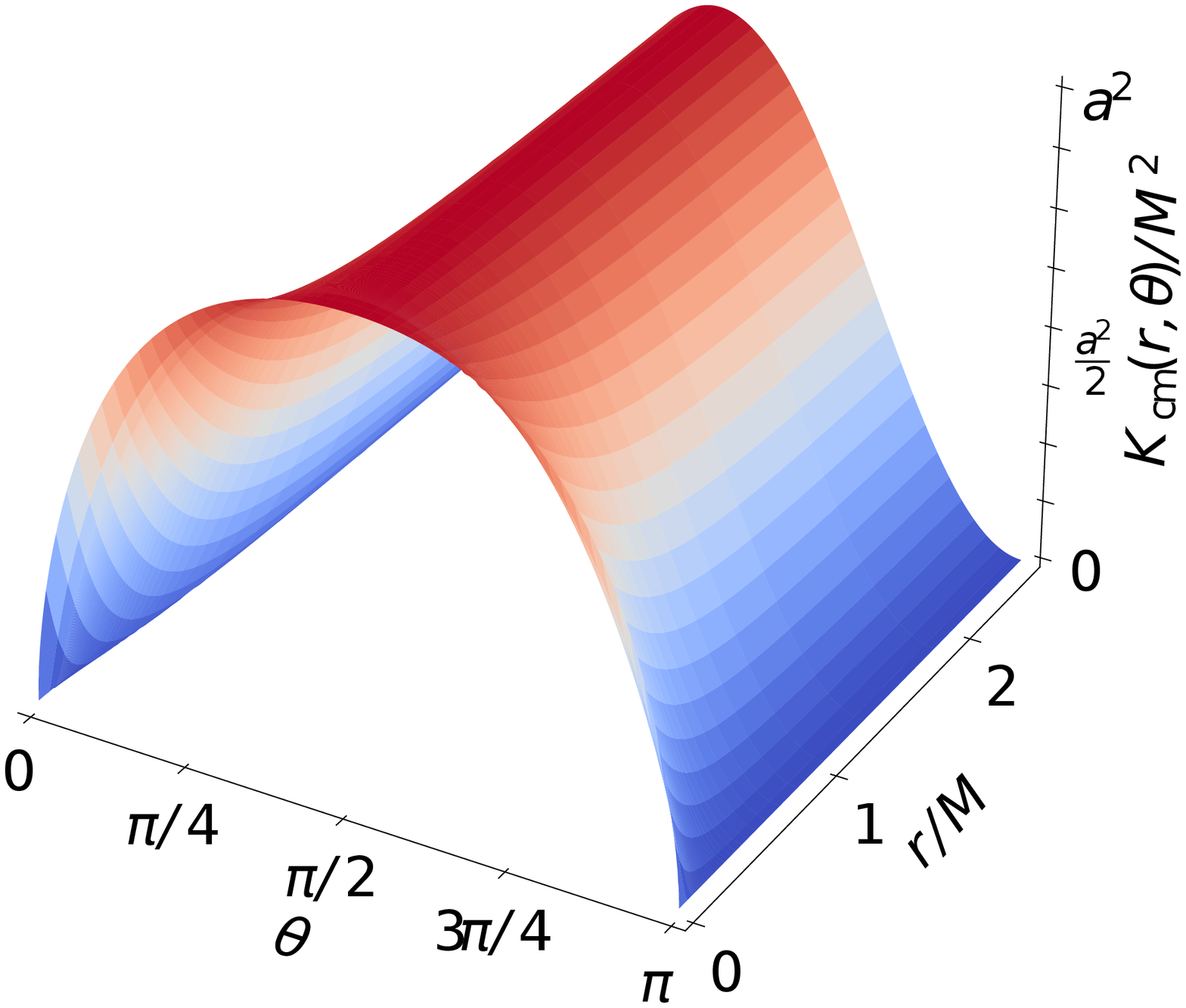}
\end{subfigure}
\caption{Function $k_{\text{cm}}(r,\theta)$ and Carter's function $K_{\text{cm}}(r,\theta)=k^2_\text{cm}\left(r,\theta\right)+a^2\sin^2\theta$ for $a=0.5M$ as a result of numerical integration of Eq.~(\ref{eq:pde_appendix}).}
\label{fig:7}
\end{figure}
\begin{figure}
\begin{subfigure}{9.2cm}
\includegraphics[width=\linewidth]{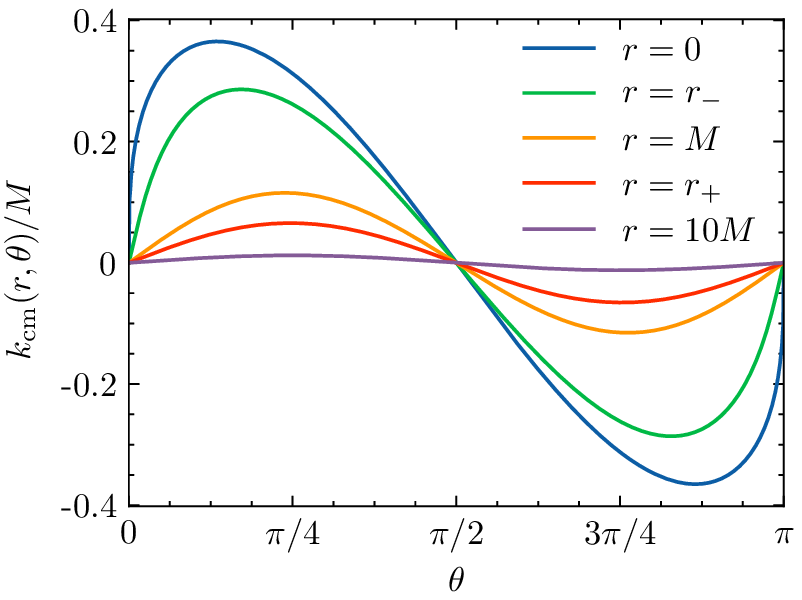}
\end{subfigure}
\begin{subfigure}{8.6cm}
\includegraphics[width=\linewidth]{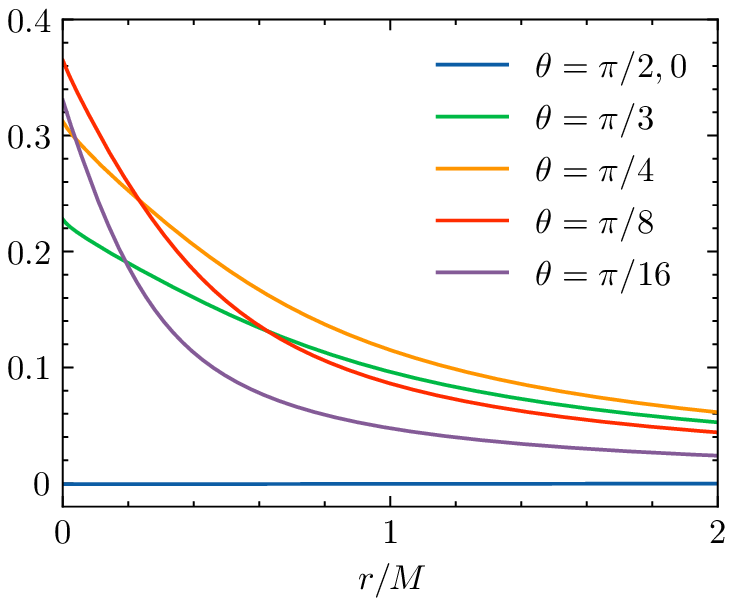}
\end{subfigure}
\caption{Function $k_\text{cm}(r,\theta)$ for $a=0.5M$, for fixed values of $r$ as a function of $\theta$ (left panel), and for fixed values of $\theta$ as a function of $r$ (right panel).}
\label{fig:8}
\end{figure}
Note that the vanishing of $k_\text{cm}$ at $\theta=\pi/2$, or equivalently $K_\text{cm}\left(r,\theta=\pi/2\right)=a^2$, is proven in the main body of the text, see below Eq.~(\ref{eq:vanish_poles}). 
The numerical integration consistently finds this property. 
In addition, we see that $K_\text{cm}\left(r,\theta\neq\pi/2\right)<a^2$ (right panel of Fig.~\ref{fig:7}). Therefore
\begin{equation}
K_{\text{cm}}(r,\theta)\leq a^2\quad\text{with equality if and only if }\theta=\pi/2,\label{eq:property}
\end{equation}
which implies that
\begin{equation}
\Sigma - k_\text{cm}^2 \geq r^2\quad\text{with equality if and only if }\theta=\pi/2,\label{eq:property_2}
\end{equation}
and that, when $\Delta\geq 0$,
\begin{equation}
\mathcal{R}_\text{cm}\geq r^4+r^2a^2+2Mra^2\quad\text{with equality if and only if }\theta=\pi/2.\label{eq:property_3}
\end{equation}

\section{Useful plots}
\label{app:plots}
This appendix gathers figures which are not particularly physically relevant for themselves, but support a number of affirmations regarding the expansions of Kerr and conformal Kerr spacetime with center-of-mass foliation.
\begin{figure}[!htb]
\begin{subfigure}{6cm}
\includegraphics[width=\linewidth]{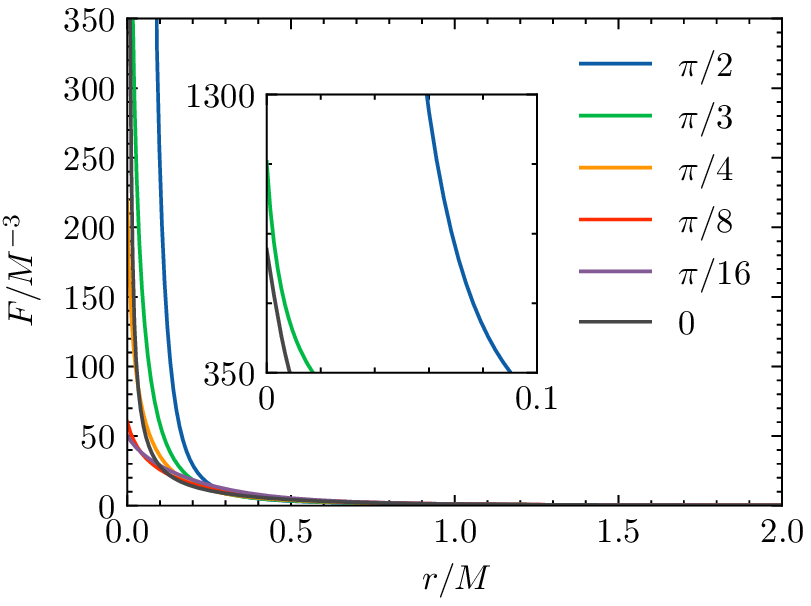}
\end{subfigure}
\begin{subfigure}{5.8cm}
\includegraphics[width=\linewidth]{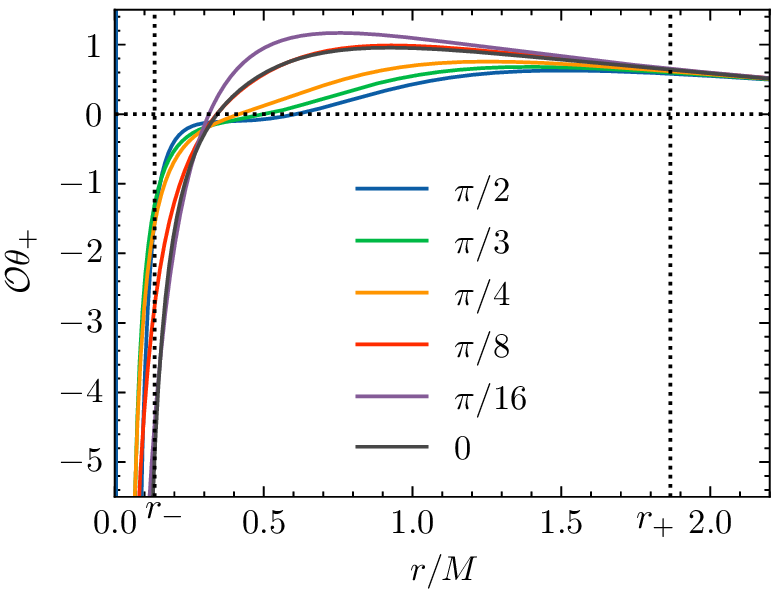}
\end{subfigure}
\begin{subfigure}{5.9cm}
\includegraphics[width=\linewidth]{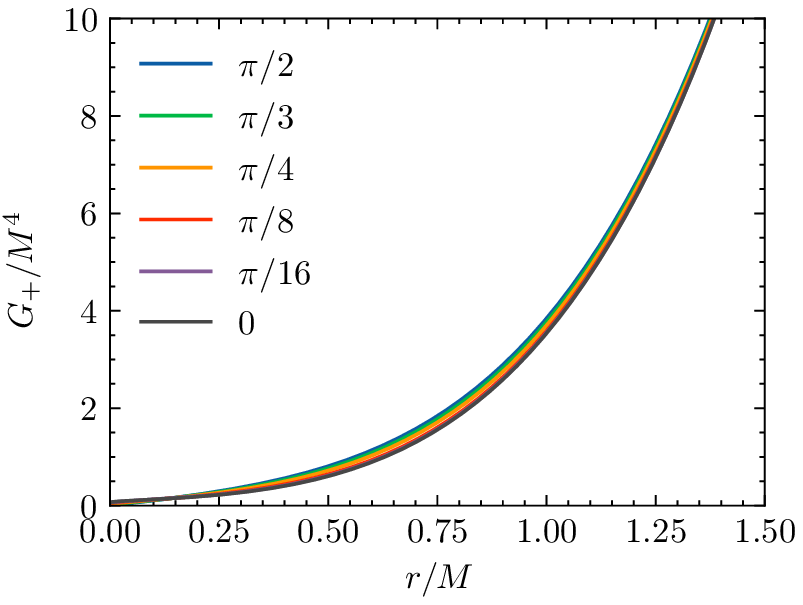}
\end{subfigure}
\caption{Three plots as functions of $r/M$ and for different values of angle $\theta$ indicated by the colors, for the Kerr metric with $a=0.5M$. Left plot: the function $F$ of Eq.~(\ref{eq:F}) for $k=k_\text{cm}$ and $K=K_\text{cm}$ is seen to be positive, and the zoom shows that it diverges at $r=0$ only for $\theta=\pi/2$. Middle plot: the quantity $\mathcal{O}\theta_+$ of Eqs.~(\ref{eq:lmoinsthetaplus}-\ref{eq:op_o}) is seen to be positive at $r_+$ and negative at $r_-$, which are indicated by dotted lines. Right plot: the function $G_+$ of Eq.~(\ref{eq:gpm}) is seen to be positive.}
\label{fig:3}
\end{figure}
\begin{figure}[!htb]
\begin{subfigure}{6.4cm}
\includegraphics[width=\linewidth]{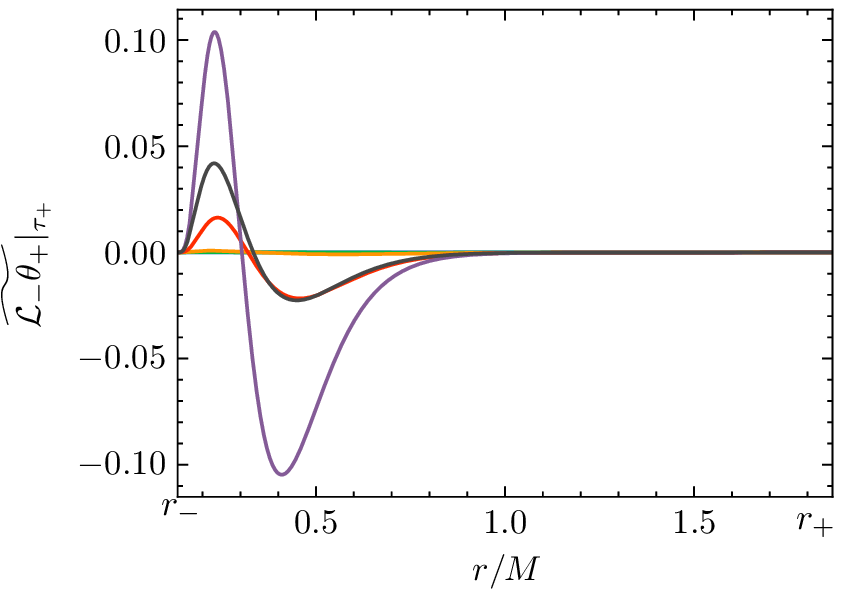}
\end{subfigure}
\begin{subfigure}{5.5cm}
\includegraphics[width=\linewidth]{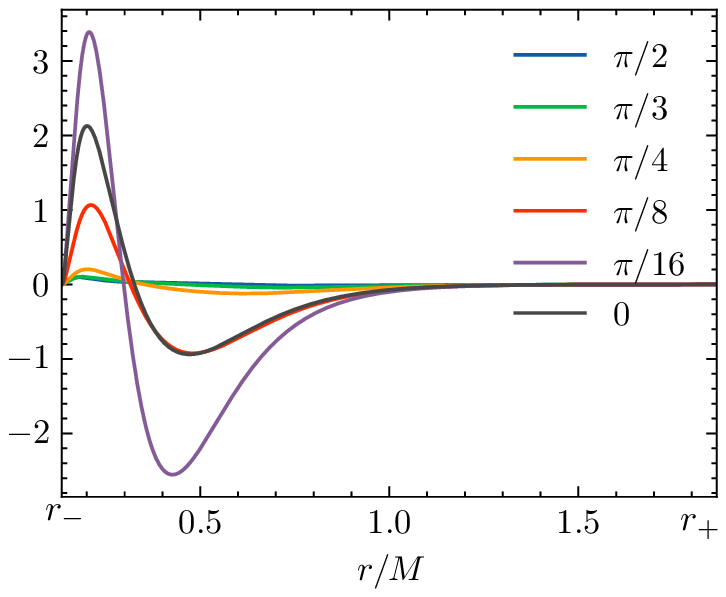}
\end{subfigure}
\begin{subfigure}{5.8cm}
\includegraphics[width=\linewidth]{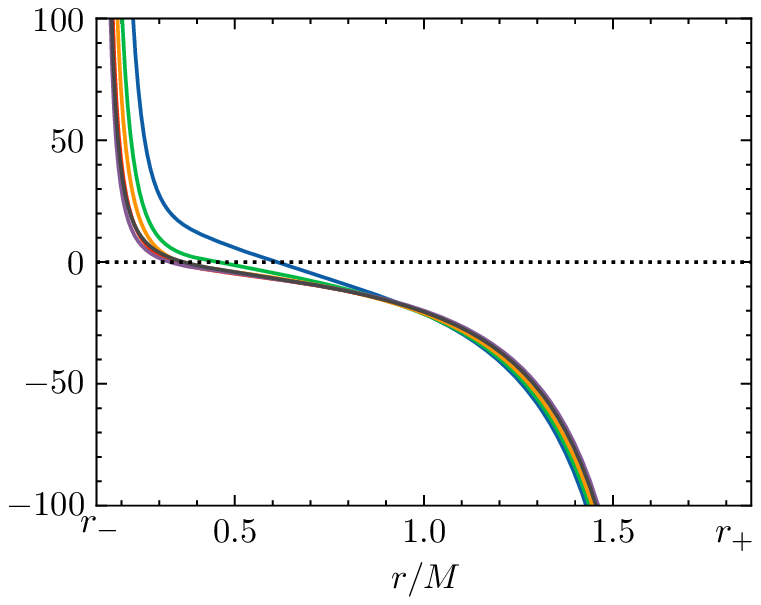}
\end{subfigure}
\caption{$\widetilde{\mathcal{L}_-\theta_+}\rvert_{\tau_+}$ for different values of angle $\theta$ indicated by the colors on the middle plot, for the conformal Kerr spacetime with $a=0.5M$, for $\alpha=2$, $1$ or $-1$ from left to right. The plot is restricted between $r_-$ and $r_+$ precisely because the trapping horizon $\tau=\tau_+$ lies within $r_-<r<r_+$. The magnitudes for $\theta=\pi/2$, $\pi/3$ and $\pi/4$ are barely distinguishable on the left and middle plots, but analysis of the values show that they obey the same pattern as all curves: positive from $r_-$ to some $r_m(\theta)$, then negative from $r_m(\theta)$ to $r_+$.}
\label{fig:6}
\end{figure} 


\end{document}